\documentclass[twocolumn]{aastex631}
\usepackage{CJK}
\usepackage{amsfonts}
\usepackage{amsmath}
\usepackage{mathrsfs}
\usepackage{epsfig}
\usepackage{subfigure}

\newcommand\diff{\,\mathrm{d}}

\newcommand\TESS{{\it TESS }}

\def\dPdt{\,(1/P_0)(\diff P_0/\diff t)}
\def\yr{\,\mathrm{yr}}
\def\cd{\,\mathrm{c}\ \mathrm{d}^{-1}}

\def\raa{Res.\ Astron.\ Astrophys.\ }

\shorttitle{Uncorrelated Variation of the Harmonics in XX Cyg}
\shortauthors{J.-S. Niu, Y. Liu, and H.-F. Xue}

\begin{document}
\begin{CJK*}{UTF8}{gbsn}

\title{Uncorrelated Amplitude and Frequency Variations of the Harmonics in SX Phoenicis Star XX Cygni}

\correspondingauthor{Jia-Shu Niu}
\email{jsniu@sxu.edu.cn}

\author[0000-0001-5232-9500]{Jia-Shu Niu (牛家树)}
\affil{Institute of Theoretical Physics, Shanxi University, Taiyuan 030006, China}
\affil{State Key Laboratory of Quantum Optics and Quantum Optics Devices, Shanxi University, Taiyuan 030006, China}

\author{Yue Liu (刘越)}
\affil{Institute of Theoretical Physics, Shanxi University, Taiyuan 030006, China}

\author[0000-0001-6027-4562]{Hui-Fang Xue (薛会芳)}
\affil{Department of Physics, Taiyuan Normal University, Jinzhong 030619, China}
\affil{Institute of Computational and Applied Physics, Taiyuan Normal University, Jinzhong 030619, China}

%%\affil{Collaborative Innovation Center of Extreme Optics, Shanxi University, Taiyuan, Shanxi 030006, China}

\begin{abstract}
Harmonics are quite common in pulsating stars, and they are always considered to mimic the behaviors of their independent parent pulsation modes and not taken as the key information for asteroseismology. Here, we report an SX Phoenicis star XX Cygni, whose periodogram is dominated by the fundamental frequency $f_{0} = 7.41481 \pm 0.00004\ \cd$ and its 19 harmonics. According to the analysis of the archival data from \TESS, we find that both the amplitudes and frequencies of the fundamental mode and the harmonics vary within \TESS Sectors 14-17 and 54-57, which might be caused by the contamination by neighbouring stars. What is more interesting is that, the harmonics show significantly uncorrelated amplitude and frequency variations over time. Some possible origins and interesting issues are proposed to scheme the further research of this hidden corner in current asteroseismology.
\end{abstract}

\section{Introduction}           %% first-level sections will be auto-capitalized

Harmonics are usually accompanied by the independent pulsation modes amongst pulsating stars, such as Cepheids \citep{Rathour2021}, RR Lyrae stars \citep{Kurtz2016}, $\delta$ Scuti stars \citep{Breger2014}, high-amplitude $\delta$ Scuti stars \citep{Niu2017}, SX Phoenicis stars \citep{Xue2020}, $\gamma$ Dor stars \citep{Kurtz2015}, pulsating white dwarfs \citep{Wu2001}, $\beta$ Cep stars \citep{Degroote2009}, and SPB stars \citep{Papics2017}. It is generally believed that the harmonics come from the non-sinusoidal of light curves, which indicates the non-linearity of the star's pulsation. Such as the non-linear transformation from the temperature variation to the flux variation ($F=\sigma T^{4}$), and the non-linear feedback of the stellar medium to the pulsation waves.
As a result, the harmonics are not considered as the intrinsic stellar pulsation modes \citep{Brickhill1992,Wu2001} and should mimic the behaviors of the parent pulsation modes. 
In practice, the harmonics are always removed in the pre-whitening process and not taken as the key information for asteroseismology.

$\delta$ Scuti stars are a class of short-period pulsating variable stars with periods between 15 minutes and 8 hours and the spectral classes A-F, which locate on the main sequence or post main sequence evolutionary stage at the bottom of the classical Cepheid instability strip and are self-excited by the $\kappa$ mechanism \citep{Breger2000,Handler2009,Uytterhoeven2011,Holdsworth2014}. High-amplitude $\delta$ Scuti stars (hereafter HADS) are a subclass of $\delta$ Scuti stars, who always have larger amplitudes and slower rotations. Most of the HADS show single or double radial pulsation modes \citep{Niu2013,Niu2017,Xue2018,Alton2019,Bowman2021,Daszynska2022,Alton2022a,Alton2022b}, and some of them have three radial pulsation modes \citep{V829Aql,GSC762110,GSC03144-595,Yang2021,V761Peg,Khruslov2014,Khruslov2022,Poleski2010,Niu2022} or even some non-radial pulsation modes \citep{Poretti2011}. SX Phoenicis (SX Phe) stars, a subgroup of HADS, are old Population II stars, which are characterized by high amplitudes of pulsation, low metallicity, and large spatial motion. 

%%(J20031564 +58571653)
XX Cygni (hereafter XX Cyg, $\alpha_{2000} = {20^h}{03^m}{16^s}$, $\delta_{2000} = 58^\circ57^\prime17^{\prime\prime}$, V=11.86 mag) is an Population II SX Phe star pulsating with a single radial mode \citep{Ceraski1904, Nijland1923, McNamara1980, Joner1982, Hintz1997, Szeidl1981, Kiss2000, Zhou2002, Blake2003}, which shows a continues linear period variation rate of $\dPdt = 1.19(13) \times 10^{-8}\ \yr^{-1}$ \citep{Yang2012}.
Although XX Cyg has a clean periodogram (which consists of the fundamental frequency $f_{0} = 7.4148\ \cd$ and its harmonics), the residuals in the $O-C$ diagram have deviations of about $0.001 - 0.005\ \mathrm{d}$, which is an order of magnitude larger than the uncertainties of the times of maximum light (TML) (about $0.0001\ \mathrm{d}$). These dispersions in the residuals indicate that the apparently stable single pulsation mode is actually not that stable.

These years, XX Cyg has been monitored by the Transiting Exoplanet Survey Satellite (\TESS \citep{Ricker2015}), whose continues photometric data provide us an opportunity to study the stability of the pulsation mode in detail.
In the case of single-mode pulsator like XX Cyg, the instability of the pulsation mode can be reflected by the variations of the amplitudes and frequencies of the fundamental mode and its harmonics.

\section{Methods}

\begin{figure*}[!htbp]
  \centering
  \includegraphics[width=0.8\textwidth, angle=0]{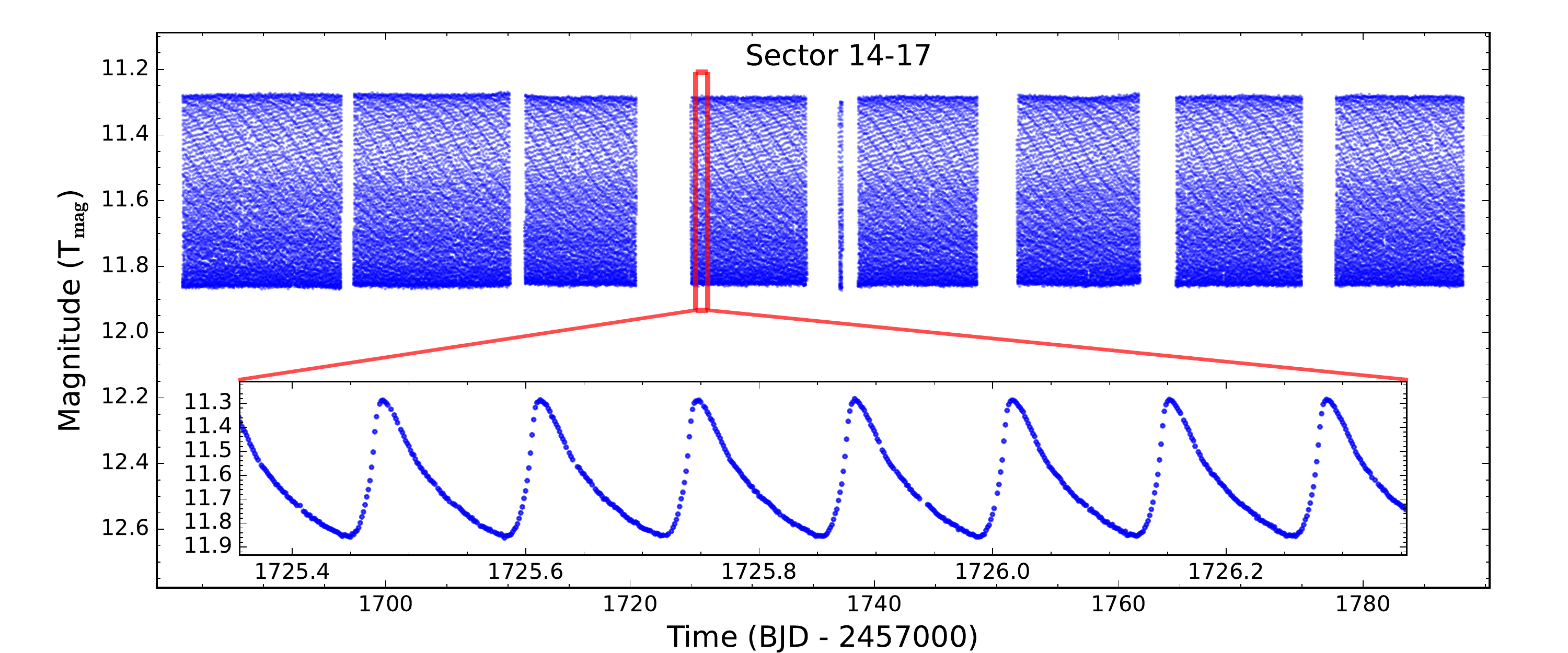}
  \includegraphics[width=0.8\textwidth, angle=0]{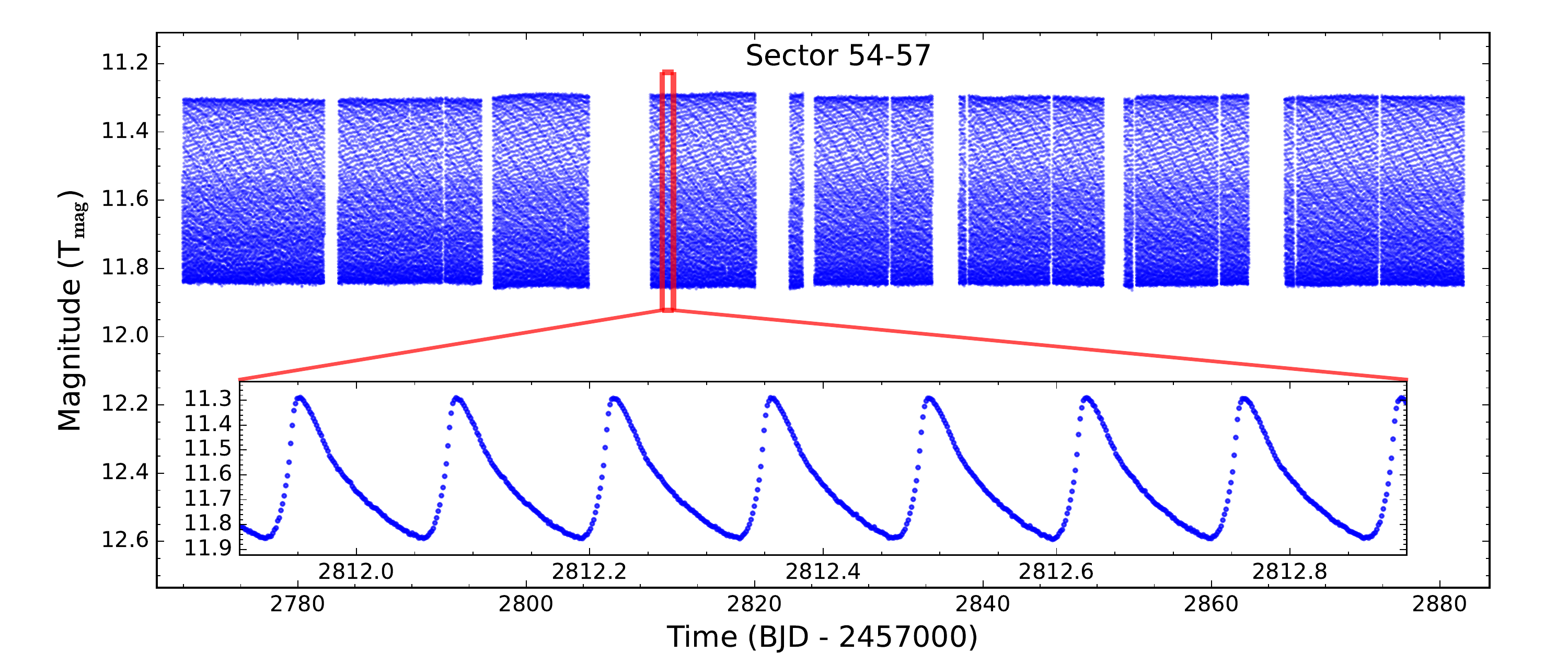}
  \caption{Light curves of XX Cyg observed by \TESS. }
  \label{fig:lc}
\end{figure*}

XX Cyg (TIC 233310793) has been observed by \TESS spacecraft continuously during Sectors 14-17 in 2019 (BJD 2458683 - 2458788, Data Set 1 (DS1)) and Sectors 54-57 in 2022 (BJD 2459770 - 2459882, Data Set 2 (DS2)), whose photometric data has gaps of about 3-4 days in the same Sector and between the continuous Sectors (see in Figure \ref{fig:lc})\footnote{XX Cyg has also been observed by \TESS in Sector 41 in 2021 (BJD 2459420 - 2459447), which is not used in this work because of the short duration of the data. }. We downloaded the 2-min cadence flux measurements which were processed by the \TESS Science Processing Operations Center (SPOC; \citet{Jenkins2016}) from MAST Portal\footnote{https://mast.stsci.edu/portal/Mashup/Clients/Mast/Portal.html}. After converting the normalized fluxes to magnitudes by utilizing the \TESS magnitude of +11.6412 and removing the long trends in each Sectors, we totally got 127055 data points (DS1+DS2). The light curves are shown in Figure \ref{fig:lc}.

\begin{figure*}[!htbp]
  \centering
  \includegraphics[width=0.8\textwidth]{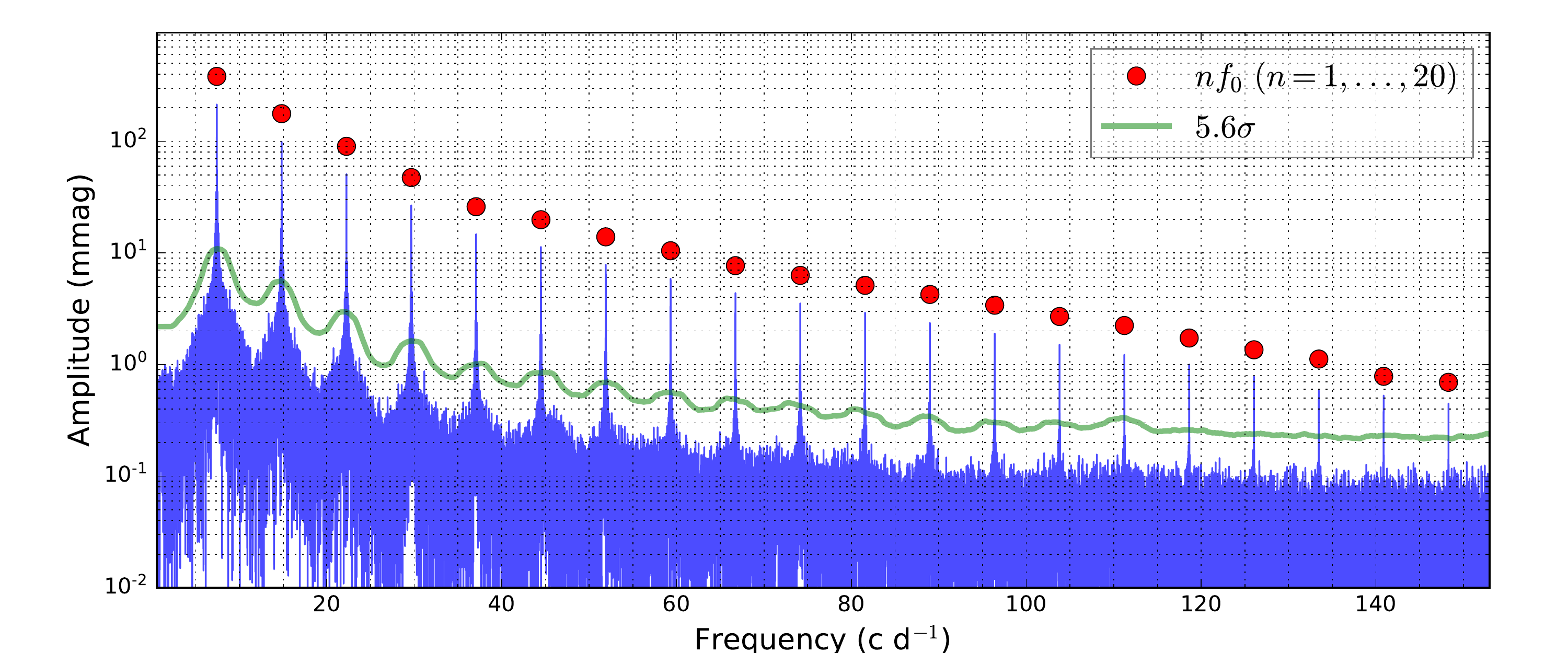}
  \caption{Periodogram of XX Cyg based on DS2. The fundamental frequency and its harmonics are marked as the red dots.}
  \label{fig:spec}
\end{figure*}

In Figure \ref{fig:spec}, the Fourier analysis was performed on DS2 to present the periodogram of XX Cyg\footnote{It is because that DS2 has more data points and longer time baseline comparing with DS1.}, in which we marked $f_{0} = 7.41481 \pm 0.00004\ \cd$ and its 19 harmonics. The detailed information of them are listed in Table \ref{tab:fre_solu}.

\begin{table*}[!htbp]
\begin{center}
  \caption{Multi-frequency solution of the \TESS light curves of XX Cyg based on DS2. Note: $\sigma_f$ denotes the error estimation of frequency, $\sigma_a$ denotes the error estimation of amplitude. S/N is calculated within a spectral window of 4 $\cd$ equally divided by the frequency peak.}
  \label{tab:fre_solu}
%%Please Capitalize the First Letter of Each Notional Word in table's caption
 \begin{tabular}{ccccccc}
   \hline
   \hline
{ID}&{Marks}&{Frequency\ $(\cd)$} &{$\sigma_f$\ $(\cd)$} &{Amplitude\ (mmag)}&{$\sigma_a$\ (mmag)} &{S/N}\\
\hline
F0 & $f_0$   &   7.41481 & 0.00004 &  214.20 & 1.94 &  110.6\\ 
F1 & $2f_0$  &  14.82961 & 0.00004 &   99.68 & 0.88 &  112.8\\  
F2 & $3f_0$  &  22.24446 & 0.00004 &   50.97 & 0.45 &  112.2\\  
F3 & $4f_0$  &  29.65926 & 0.00004 &   26.75 & 0.24 &  110.8\\  
F4 & $5f_0$  &  37.07406 & 0.00004 &   14.73 & 0.13 &  109.3\\  
F5 & $6f_0$  &  44.48886 & 0.00005 &   11.23 & 0.10 &  107.8\\  
F6 & $7f_0$  &  51.90371 & 0.00004 &    7.89 & 0.07 &  110.2\\  
F7 & $8f_0$  &  59.31851 & 0.00005 &    5.90 & 0.06 &  103.9\\  
F8 & $9f_0$  &  66.73327 & 0.00005 &    4.41 & 0.04 &   98.3\\   
F9 & $10f_0$ &  74.14811 & 0.00005 &    3.59 & 0.04 &   89.6\\   
F10& $11f_0$ &  81.56296 & 0.00006 &    2.94 & 0.03 &   86.8\\   
F11& $12f_0$ &  88.97771 & 0.00006 &    2.39 & 0.03 &   78.1\\   
F12& $13f_0$ &  96.39256 & 0.00007 &    1.94 & 0.03 &   70.7\\   
F13& $14f_0$ & 103.80741 & 0.00008 &    1.55 & 0.03 &   61.2\\  
F14& $15f_0$ & 111.22216 & 0.00009 &    1.25 & 0.02 &   52.8\\  
F15& $16f_0$ & 118.6372  & 0.0001  &    1.01 & 0.02 &   45.1\\  
F16& $17f_0$ & 126.0519  & 0.0001  &    0.78 & 0.02 &   38.1\\  
F17& $18f_0$ & 133.4666  & 0.0002  &    0.60 & 0.02 &   30.2\\  
F18& $19f_0$ & 140.8813  & 0.0002  &    0.46 & 0.02 &   24.0\\  
F19& $20f_0$ & 148.2962  & 0.0003  &    0.34 & 0.02 &   18.6\\  
   \hline
   \hline
\end{tabular}
%%\tablefoot{$\sigma_f$ denotes the error estimation of frequency, $\sigma_a$ denotes the error estimation of amplitude. S/N is calculated within a spectral window of 4 $\cd$ equally divided by the frequency peak.}
\end{center}
\end{table*}

\begin{figure*}[!htbp]
  \centering
  \includegraphics[width=0.495\textwidth]{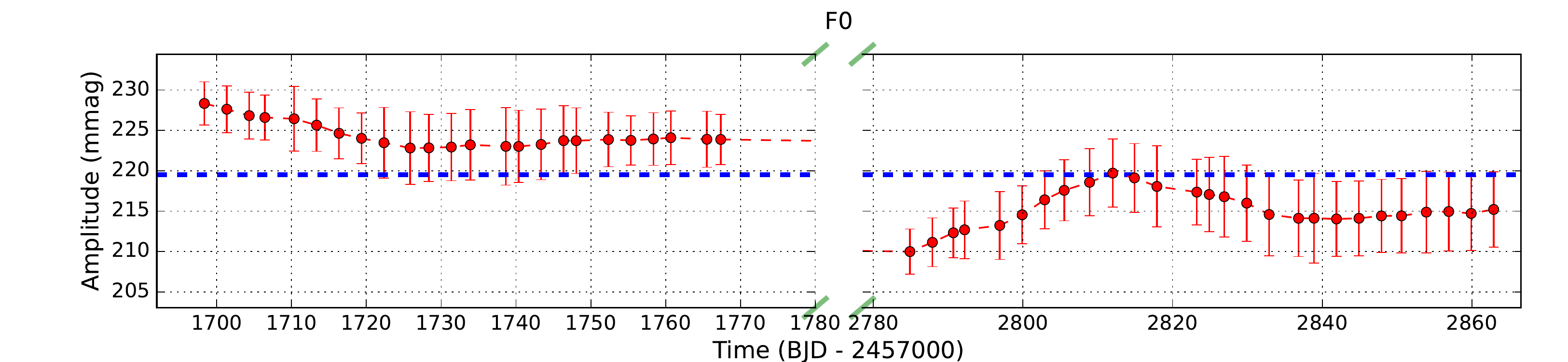}
  \includegraphics[width=0.495\textwidth]{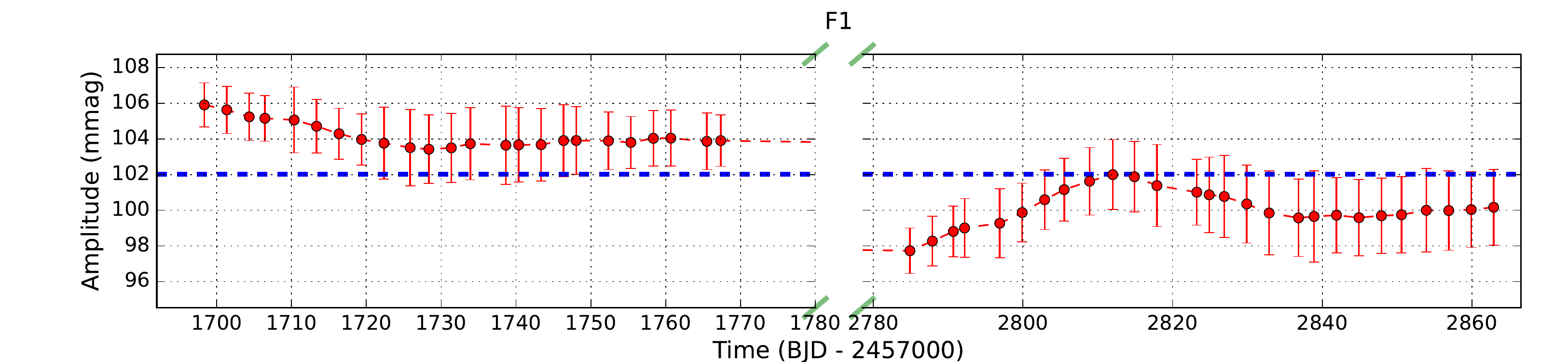}
  \includegraphics[width=0.495\textwidth]{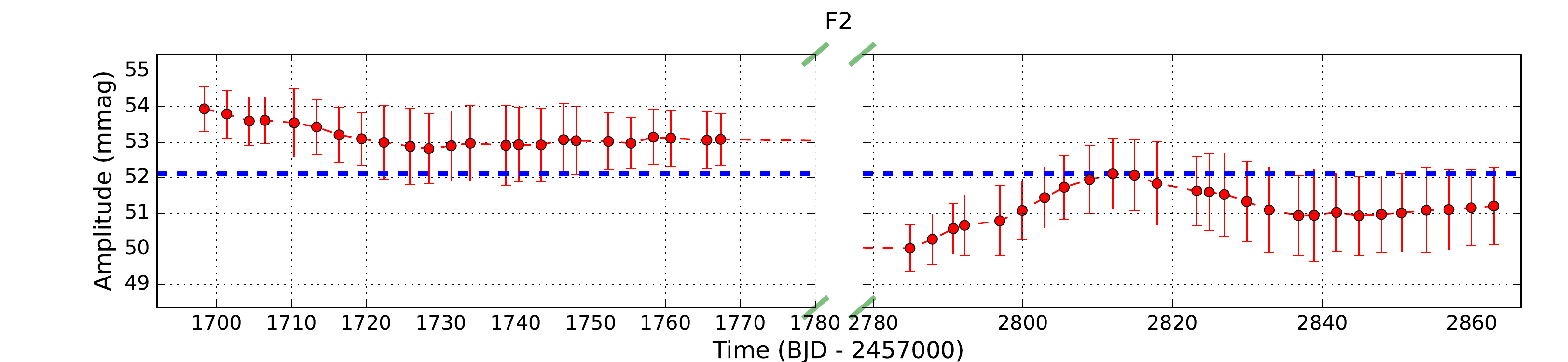}
  \includegraphics[width=0.495\textwidth]{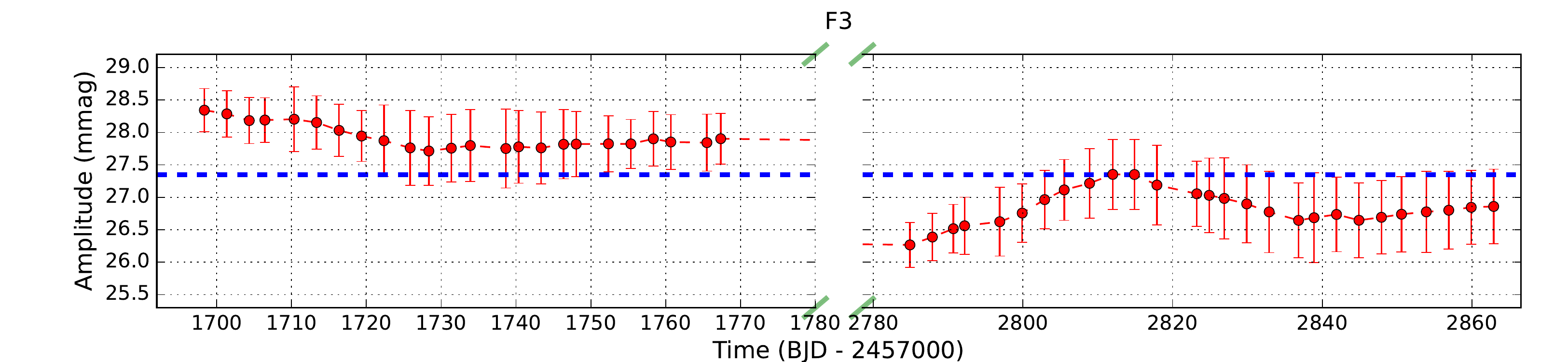}
  \includegraphics[width=0.495\textwidth]{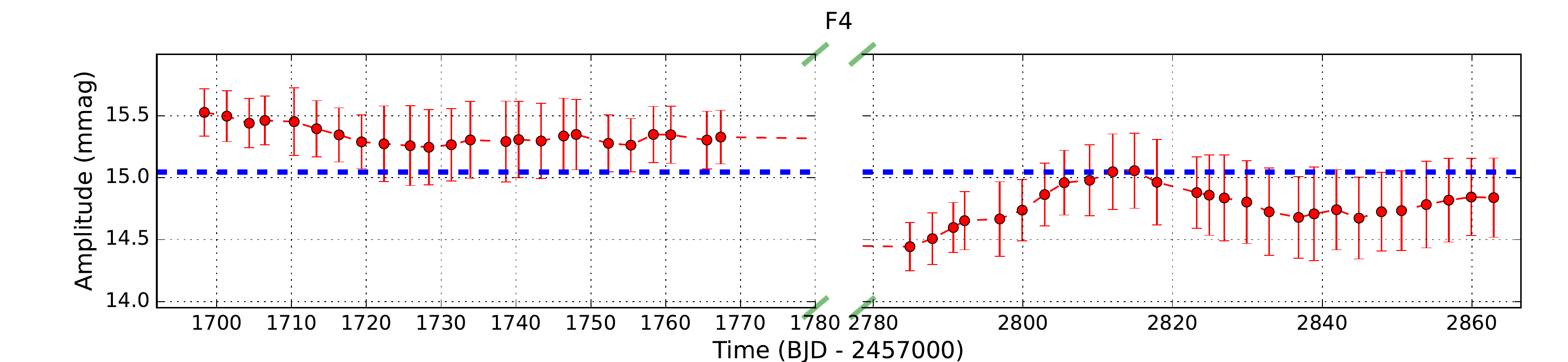}
  \includegraphics[width=0.495\textwidth]{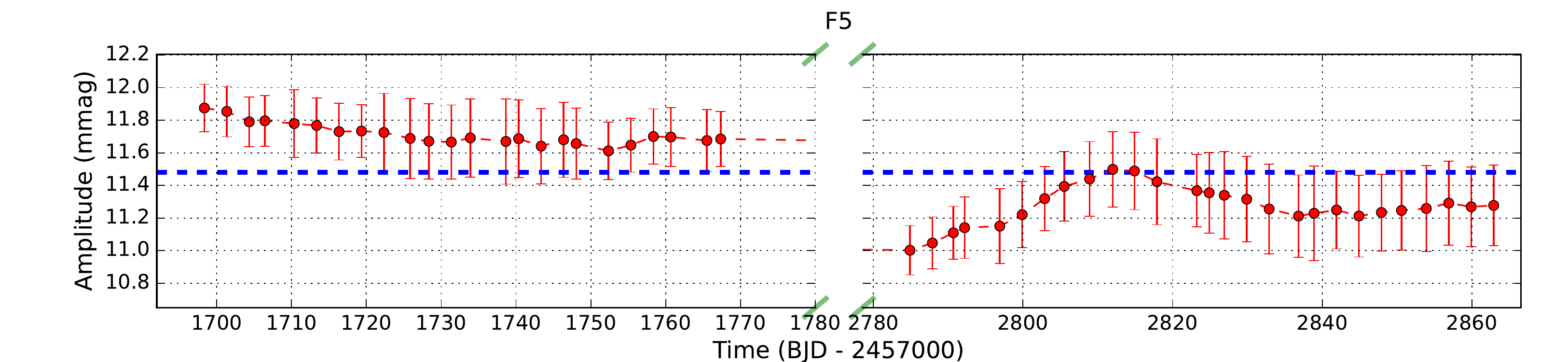}
  \includegraphics[width=0.495\textwidth]{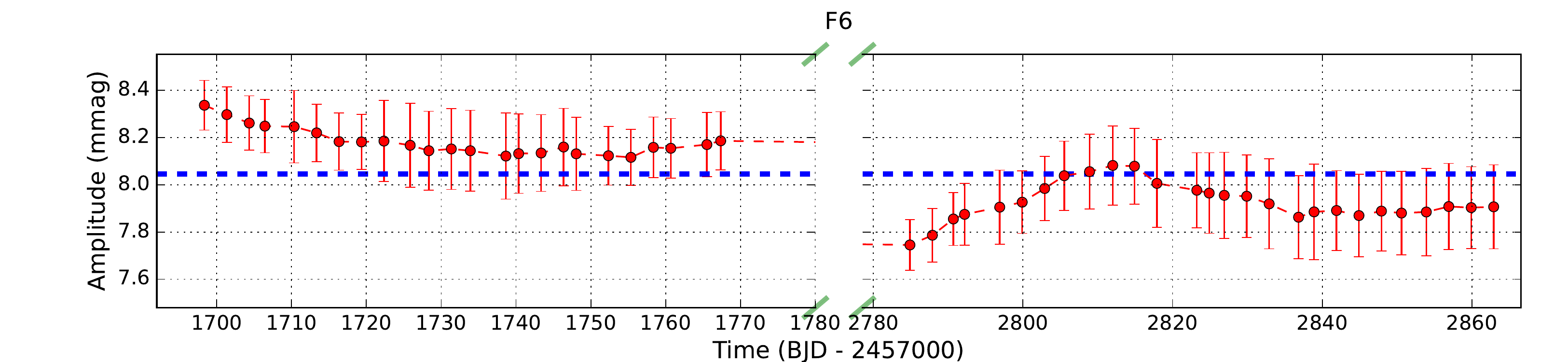}
  \includegraphics[width=0.495\textwidth]{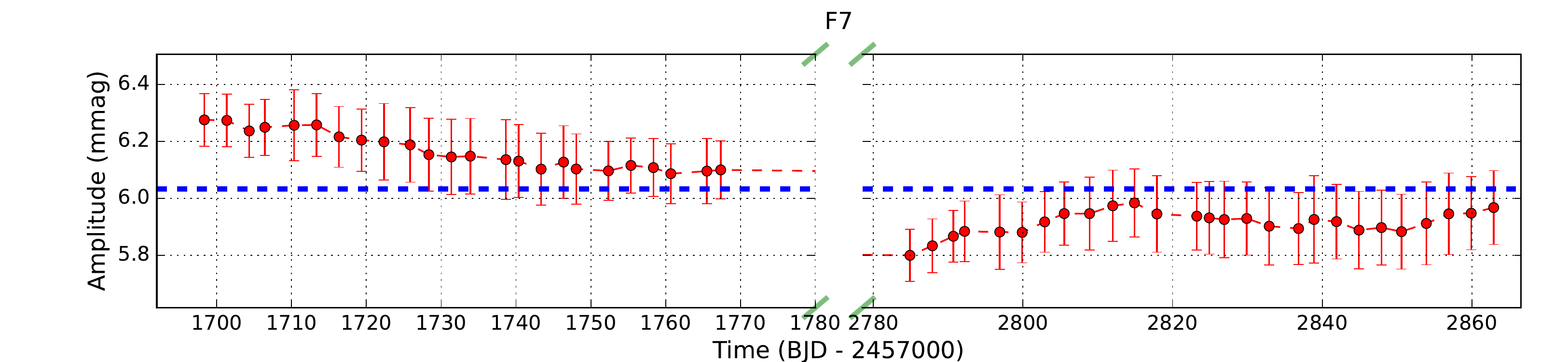}
  \includegraphics[width=0.495\textwidth]{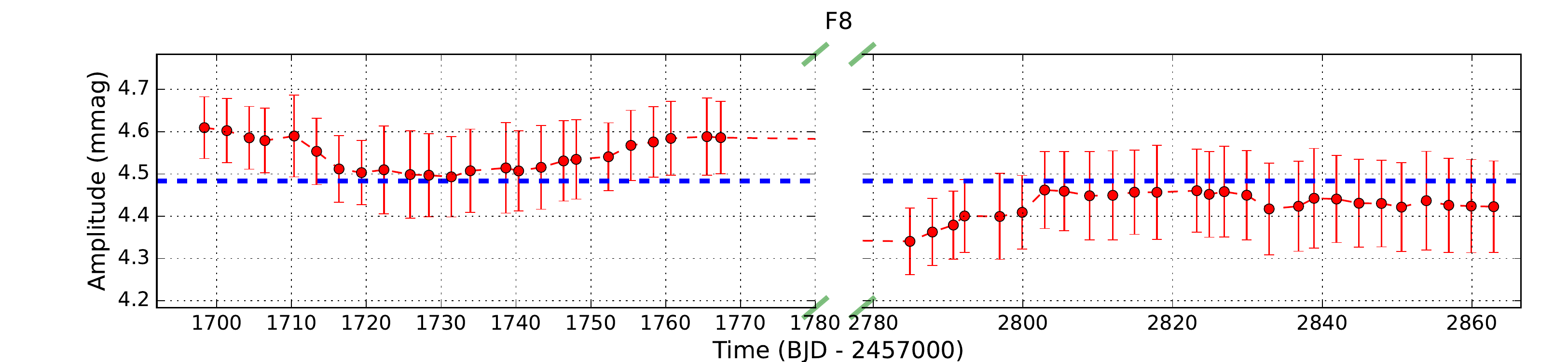}
  \includegraphics[width=0.495\textwidth]{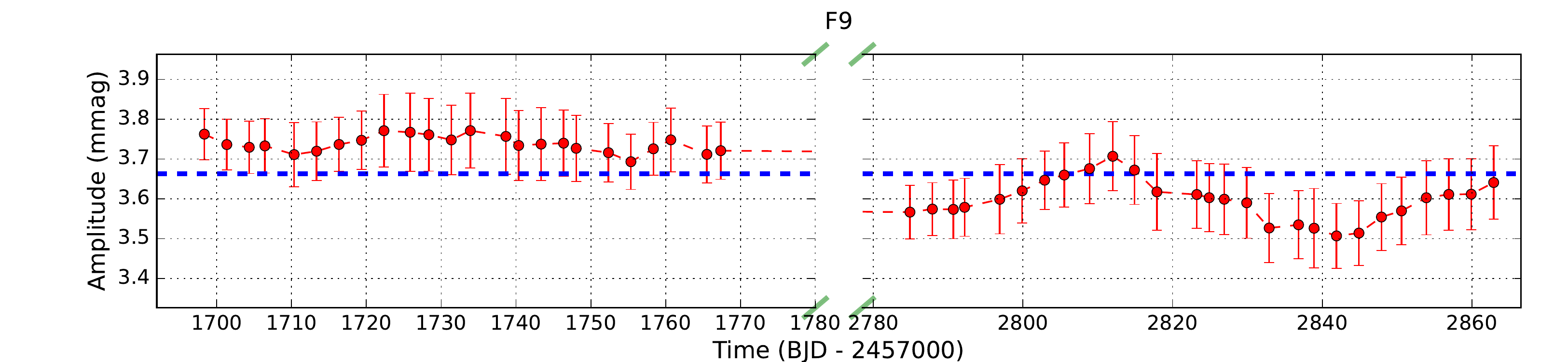}
  \includegraphics[width=0.495\textwidth]{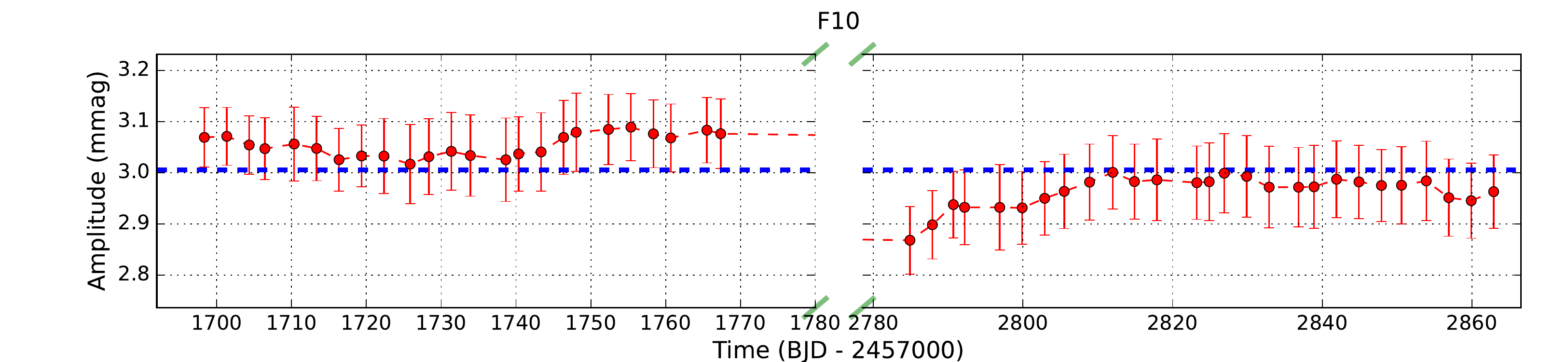}
  \includegraphics[width=0.495\textwidth]{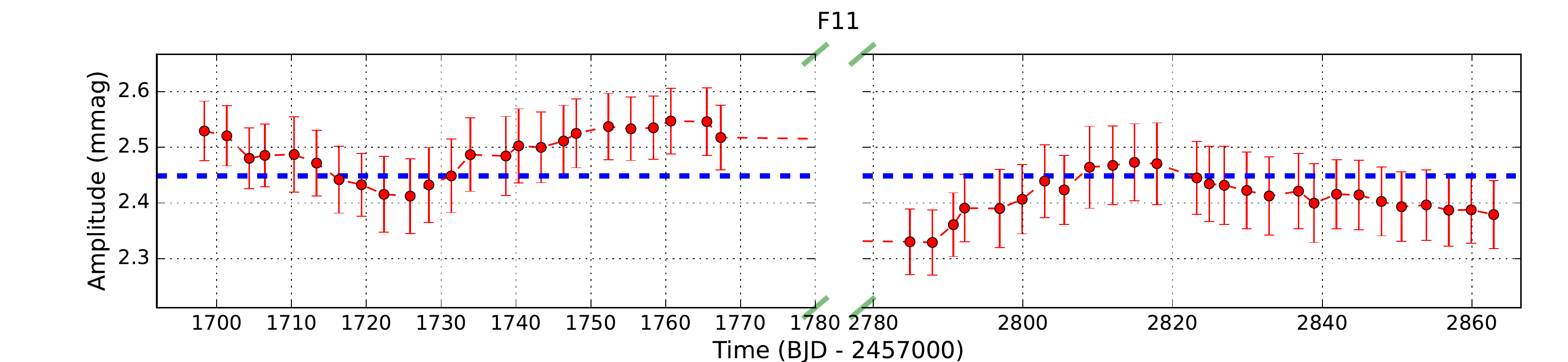}
  \includegraphics[width=0.495\textwidth]{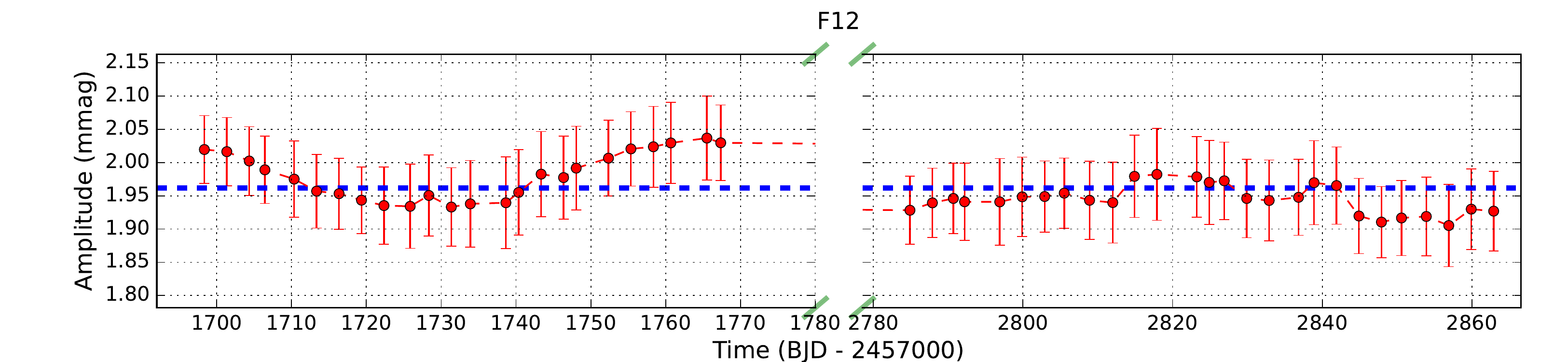}
  \includegraphics[width=0.495\textwidth]{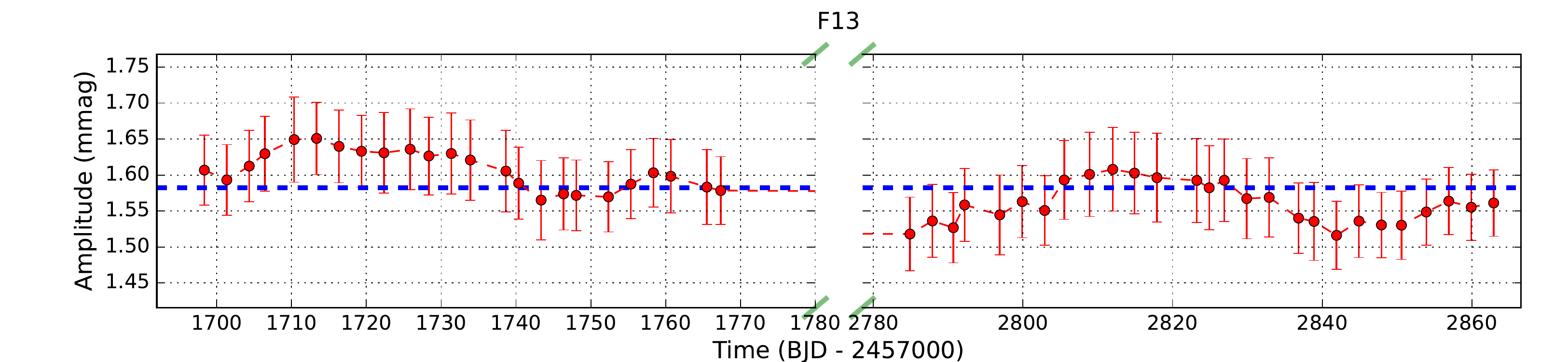}
  \includegraphics[width=0.495\textwidth]{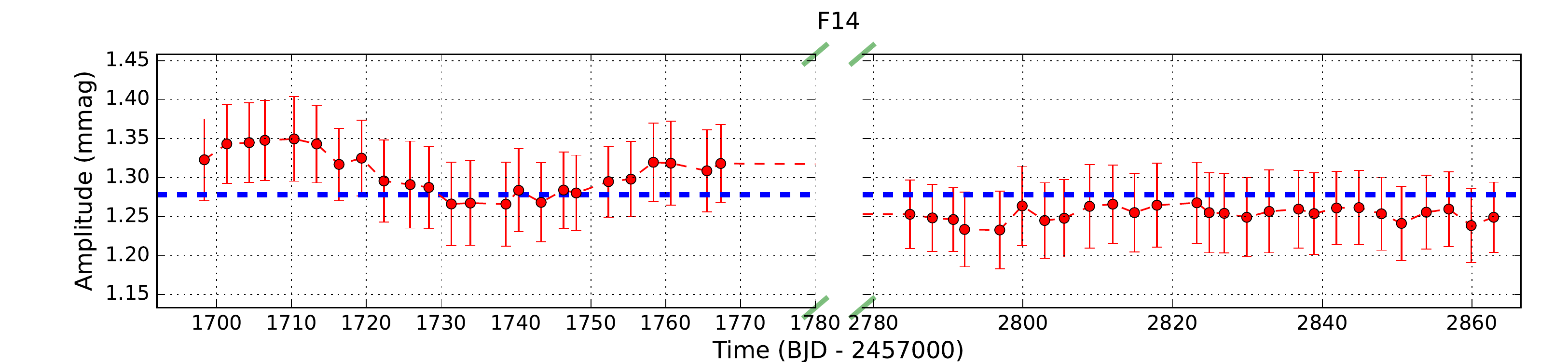}
  \includegraphics[width=0.495\textwidth]{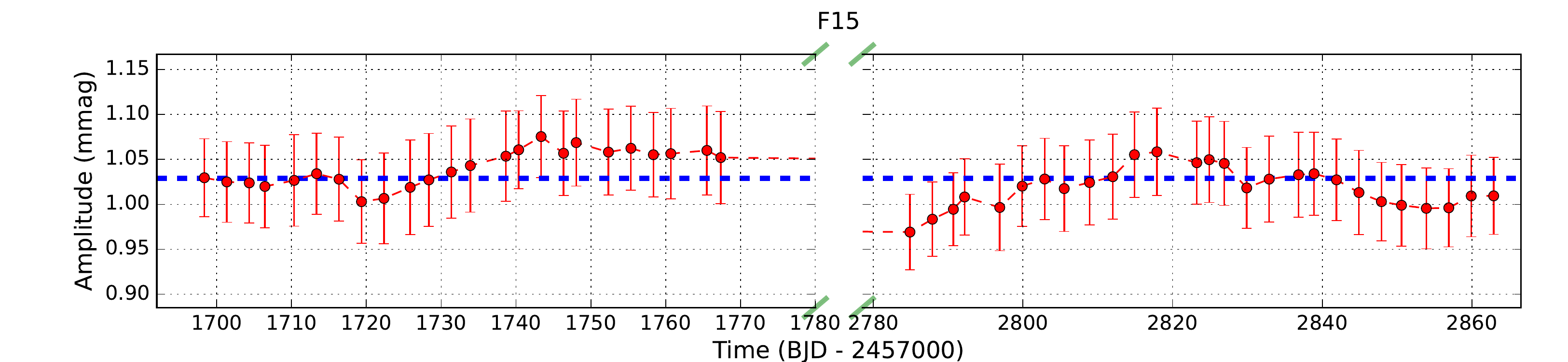}
  \includegraphics[width=0.495\textwidth]{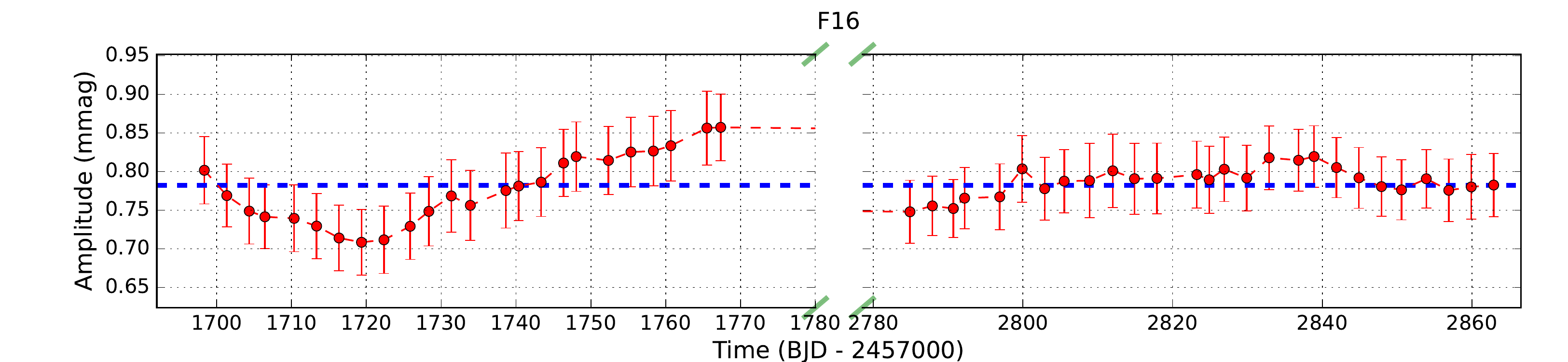}
  \includegraphics[width=0.495\textwidth]{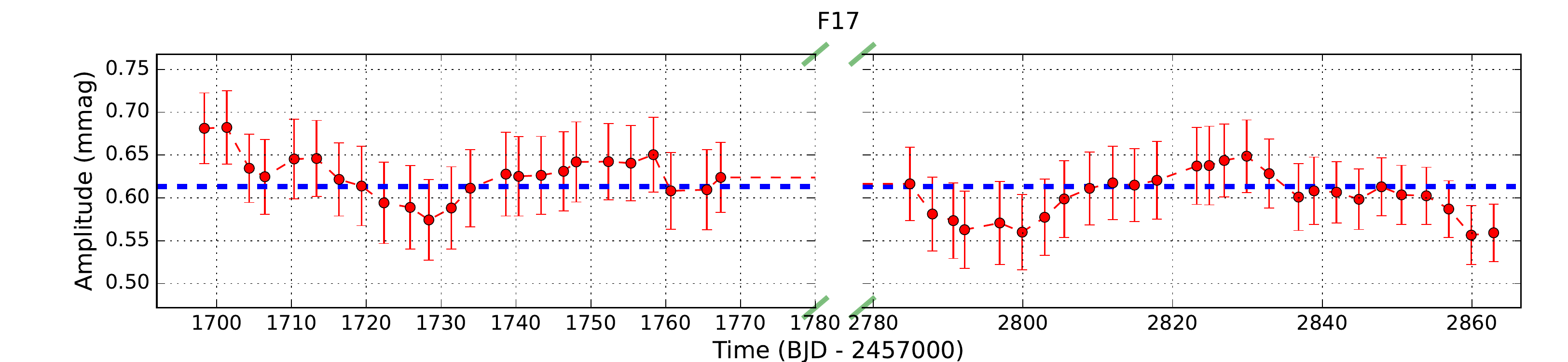}
  \includegraphics[width=0.495\textwidth]{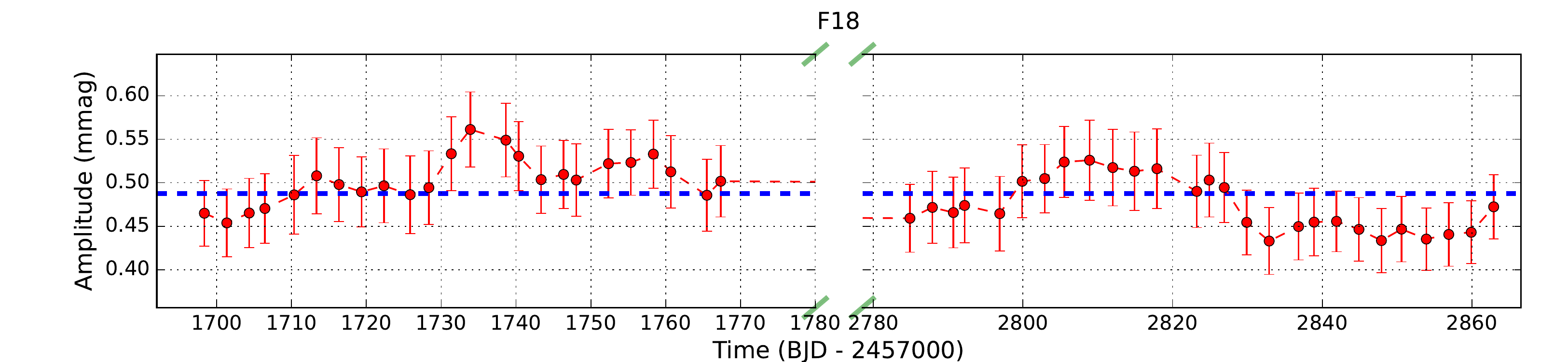}
  \includegraphics[width=0.495\textwidth]{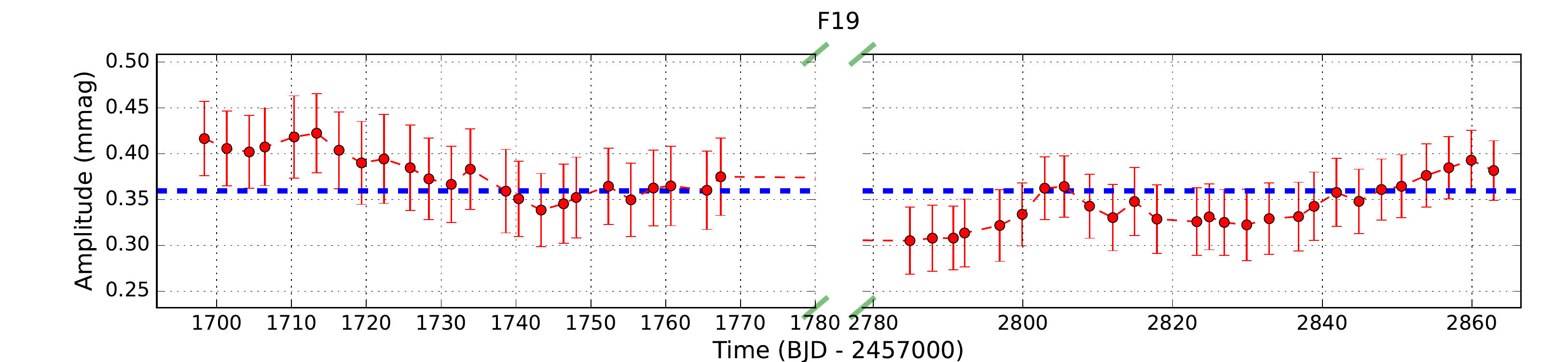}
  \caption{Variations of the amplitudes of F0 to F19. The red dots with errorbars present the amplitudes in moving windows, and the blue dashed lines present the time-averaged amplitudes.}
  \label{fig:var_amp}
\end{figure*}

\begin{figure*}[!htbp]
  \centering
  \includegraphics[width=0.495\textwidth]{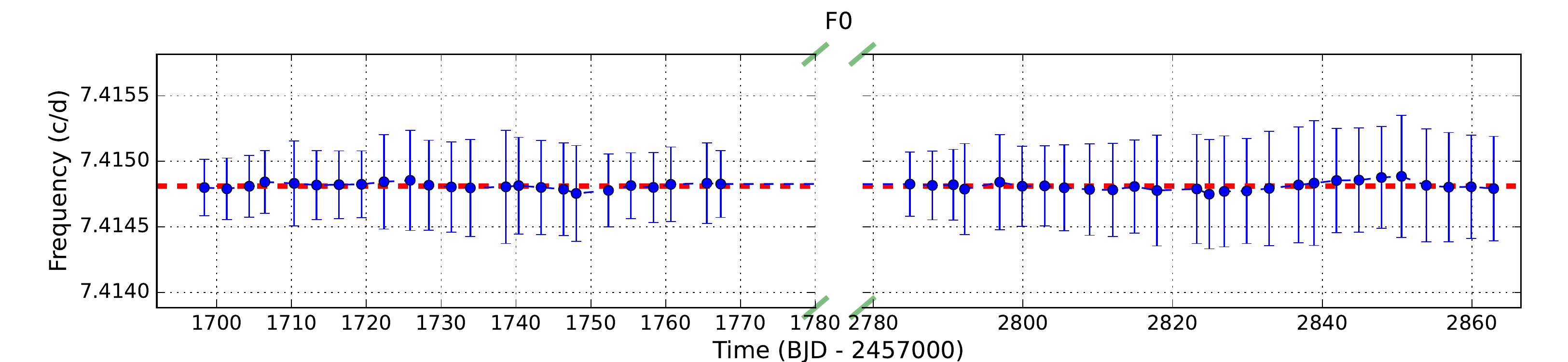}
  \includegraphics[width=0.495\textwidth]{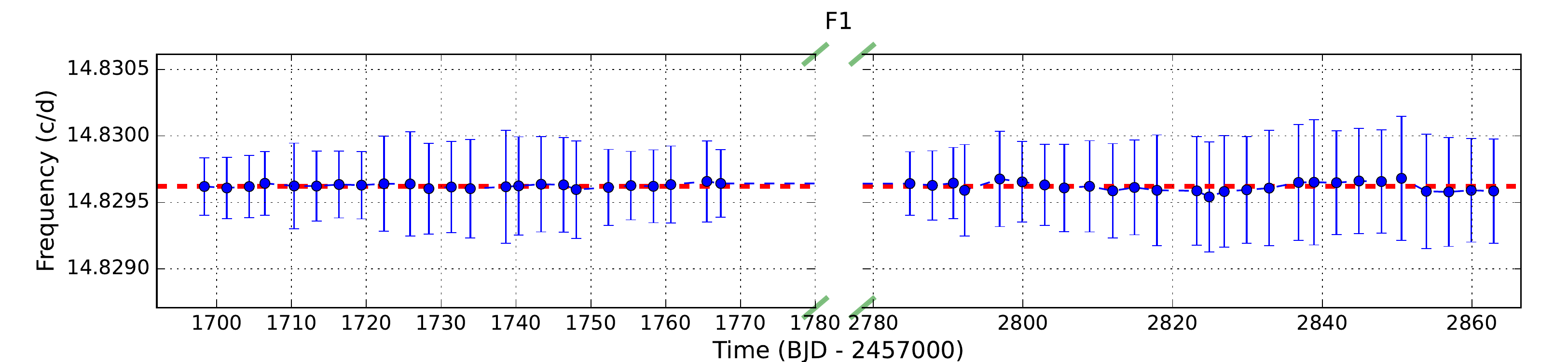}
  \includegraphics[width=0.495\textwidth]{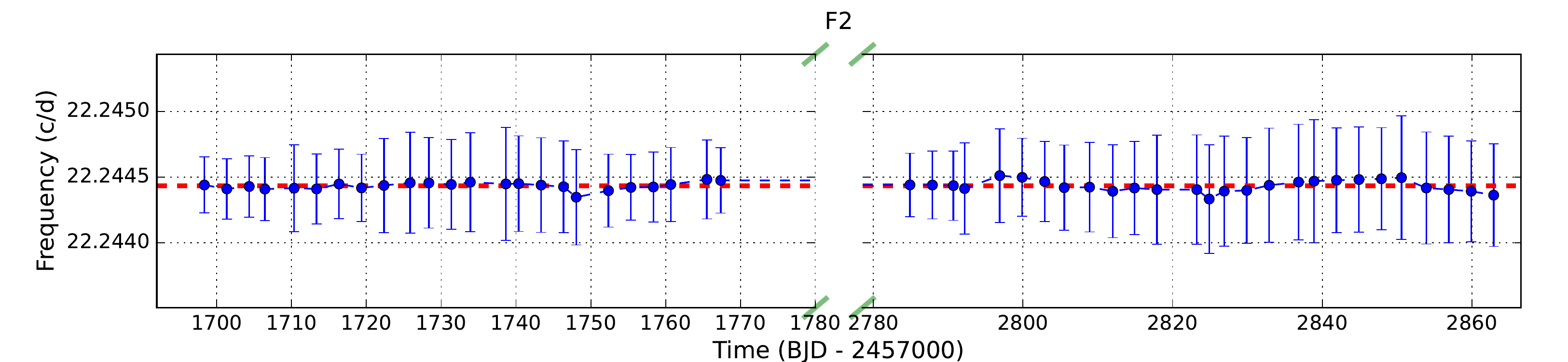}
  \includegraphics[width=0.495\textwidth]{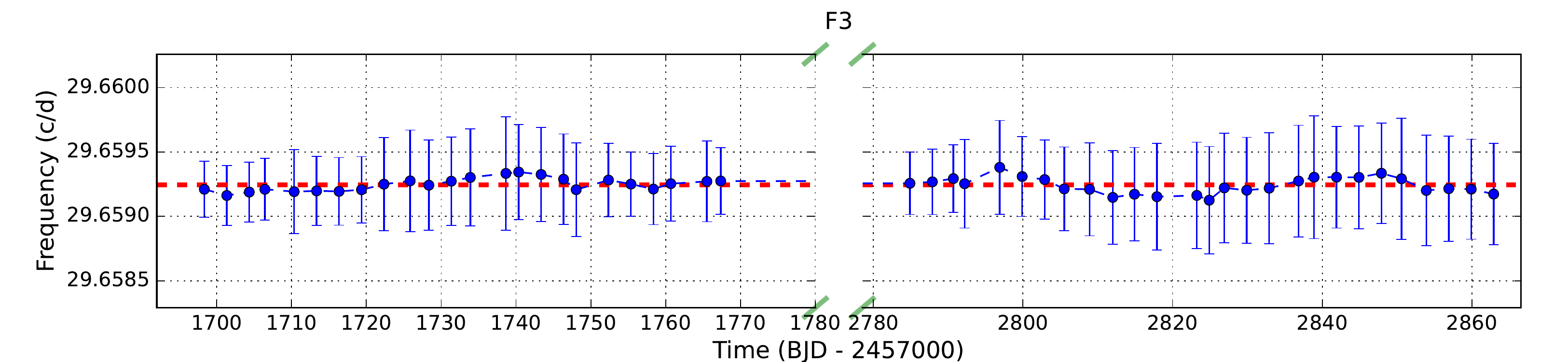}
  \includegraphics[width=0.495\textwidth]{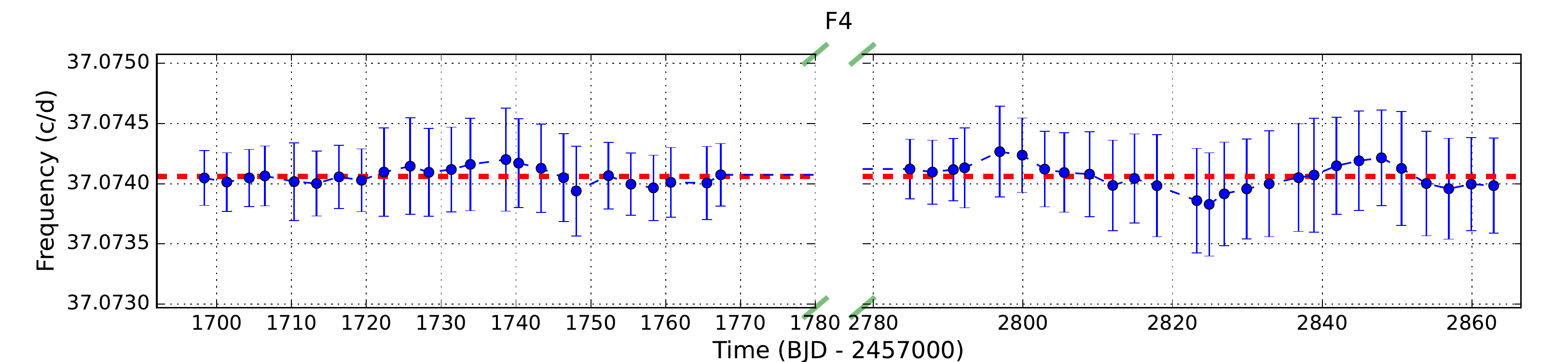}
  \includegraphics[width=0.495\textwidth]{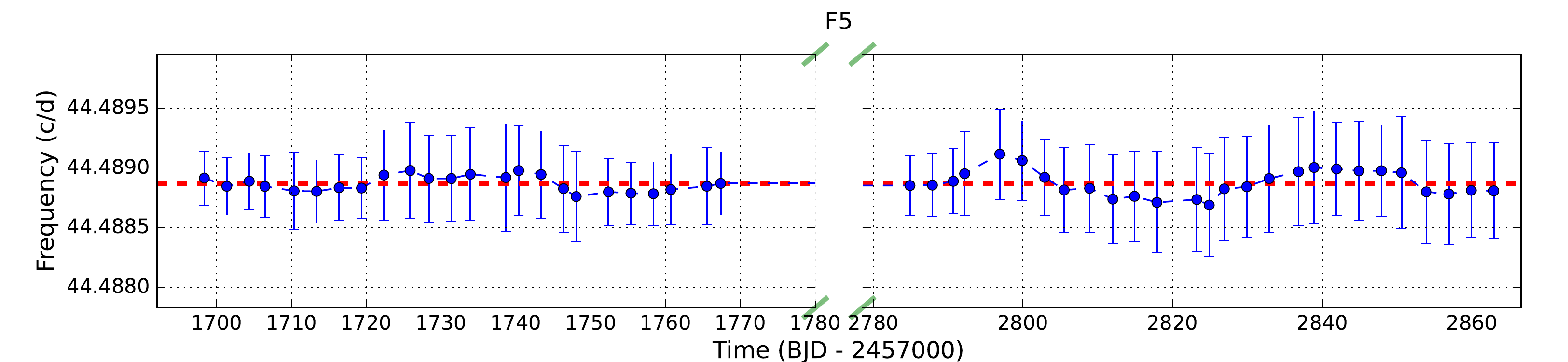}
  \includegraphics[width=0.495\textwidth]{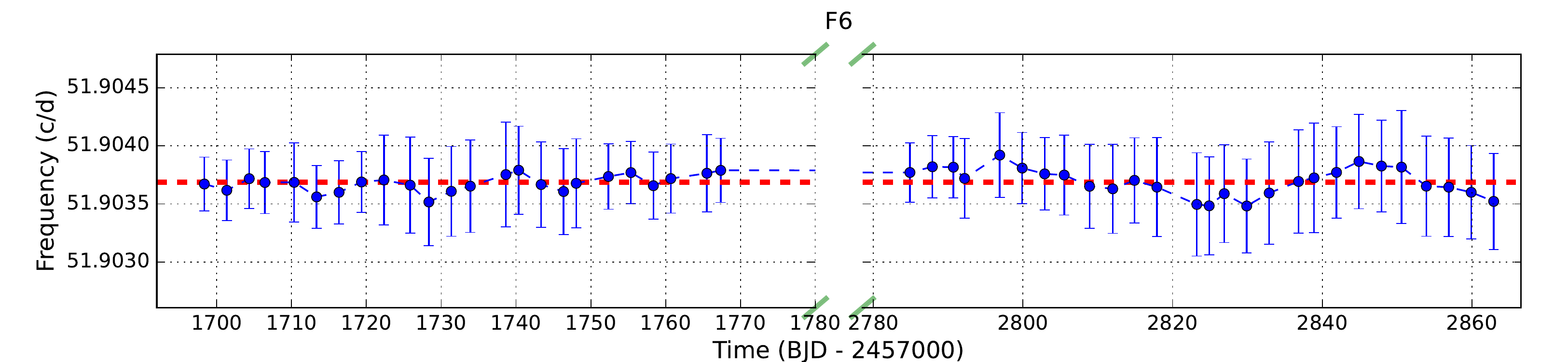}
  \includegraphics[width=0.495\textwidth]{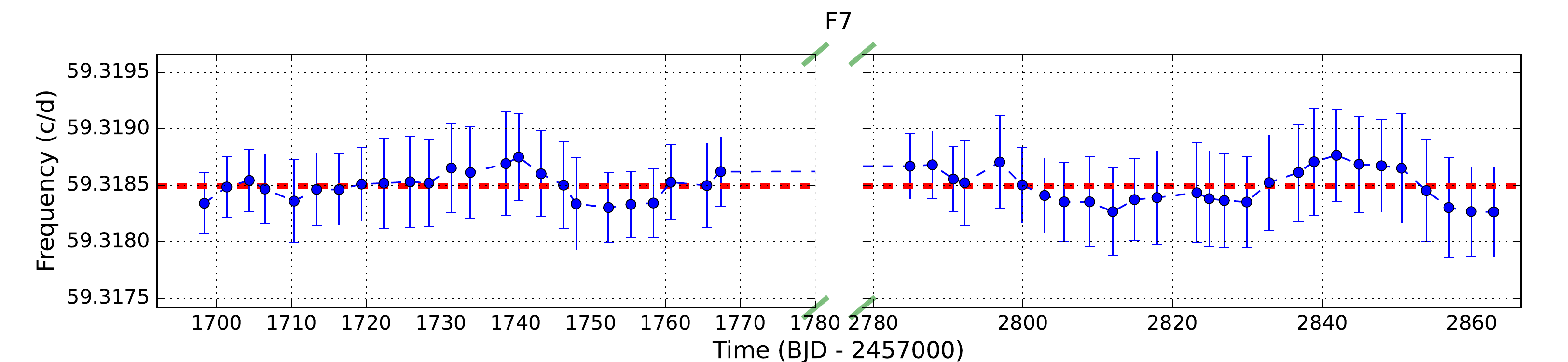}
  \includegraphics[width=0.495\textwidth]{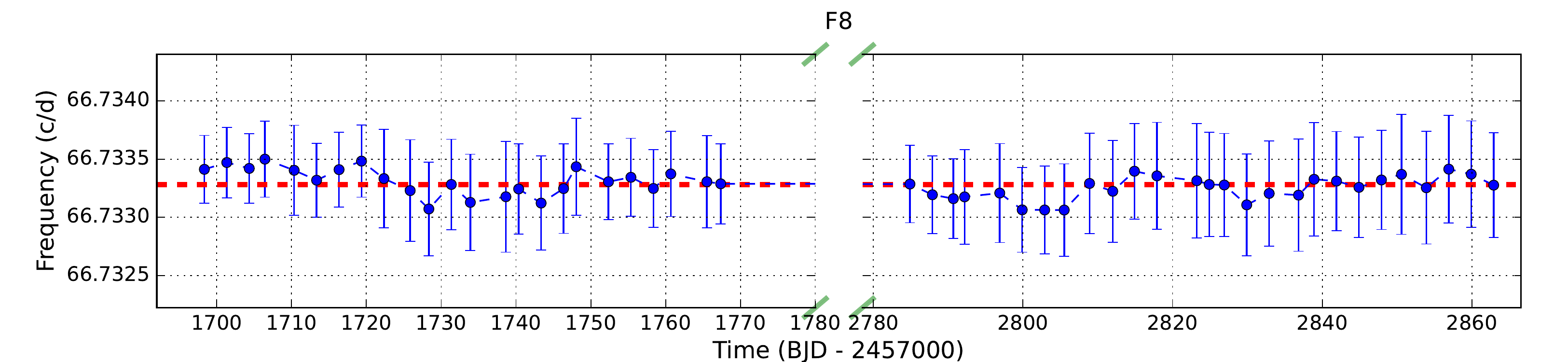}
  \includegraphics[width=0.495\textwidth]{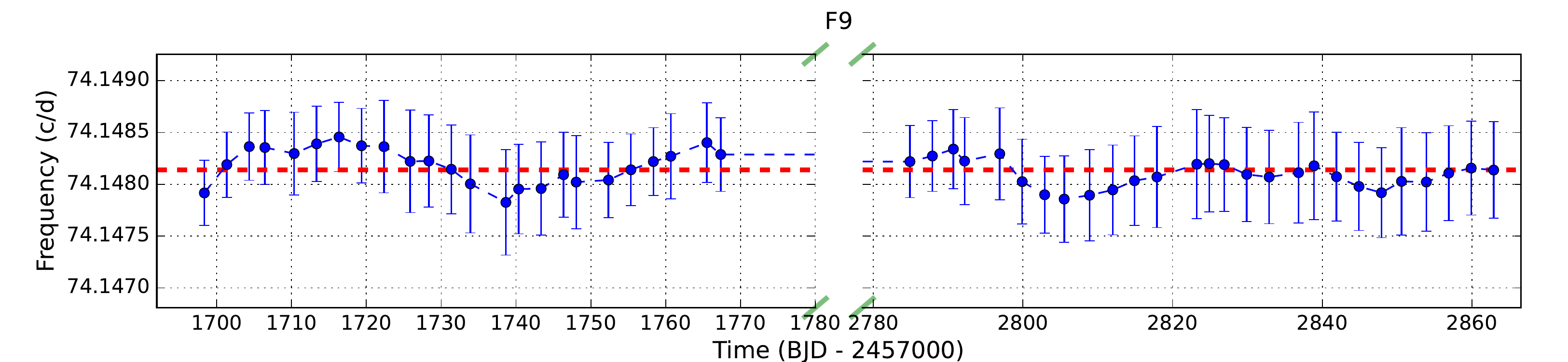}
  \includegraphics[width=0.495\textwidth]{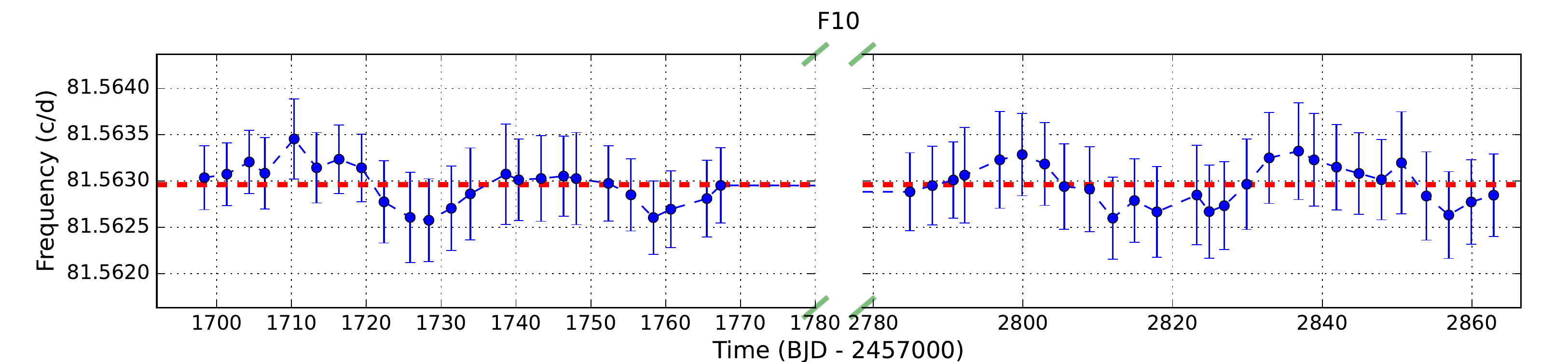}
  \includegraphics[width=0.495\textwidth]{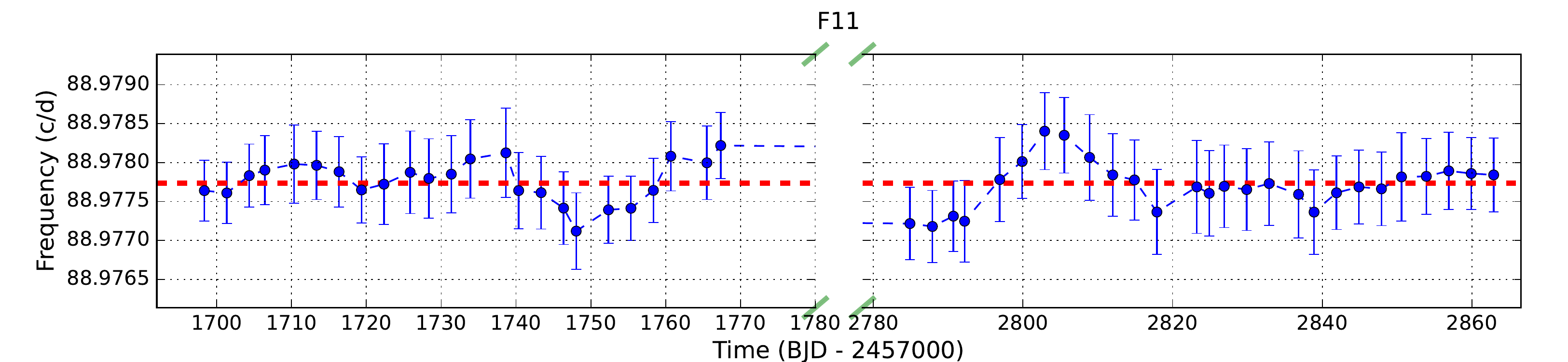}
  \includegraphics[width=0.495\textwidth]{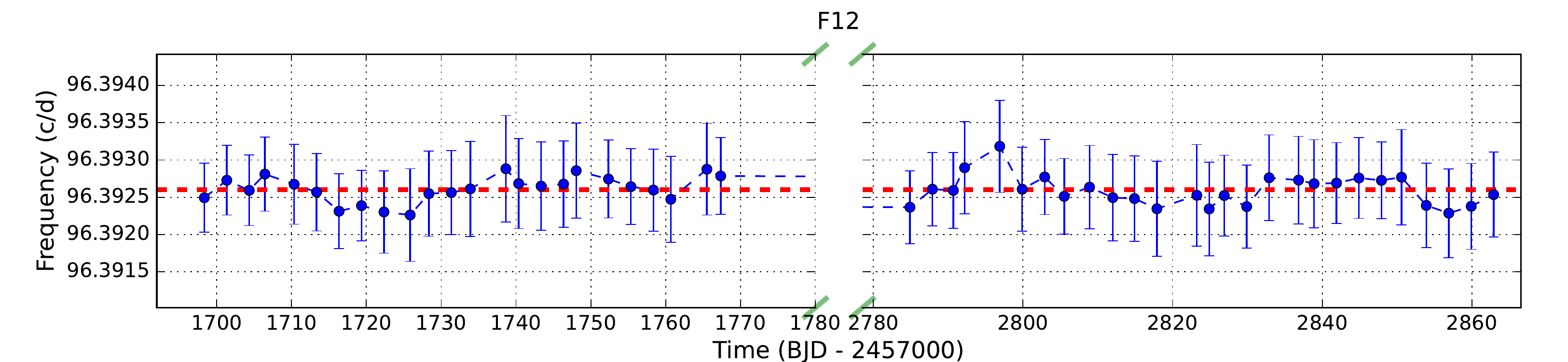}
  \includegraphics[width=0.495\textwidth]{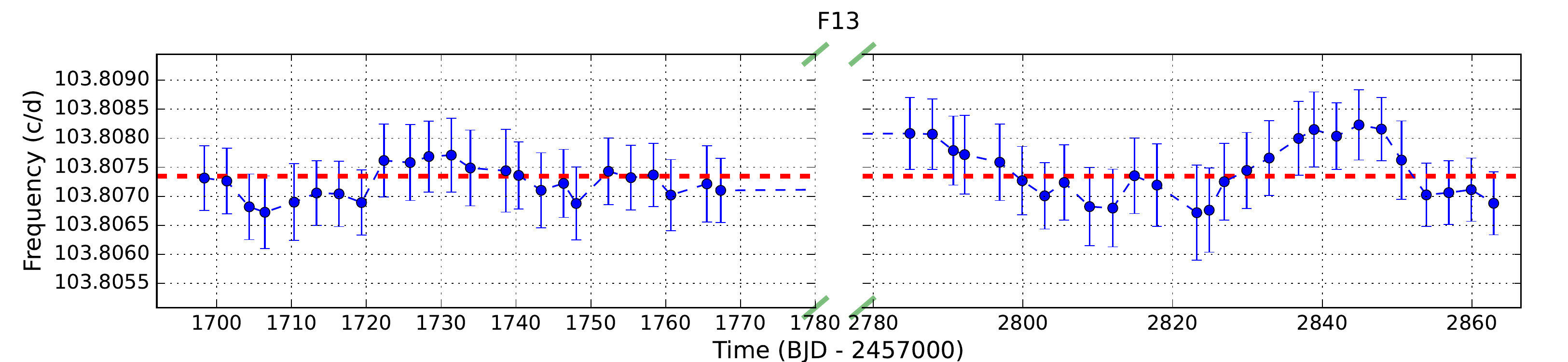}
  \includegraphics[width=0.495\textwidth]{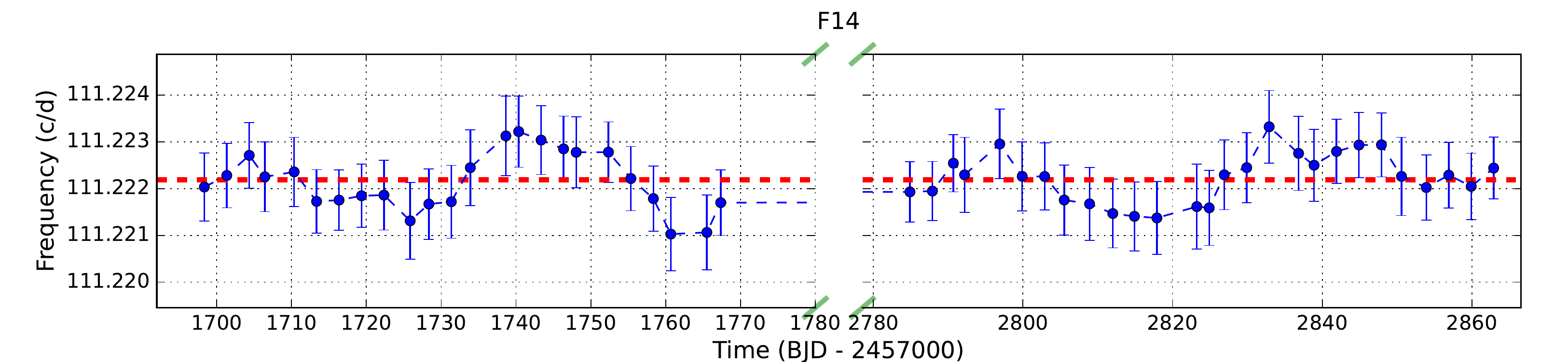}
  \includegraphics[width=0.495\textwidth]{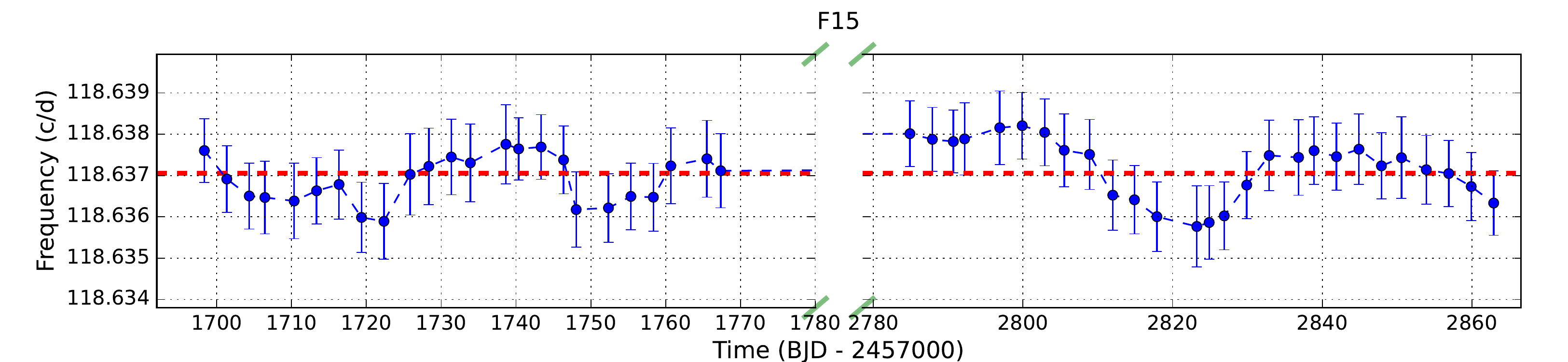}
  \includegraphics[width=0.495\textwidth]{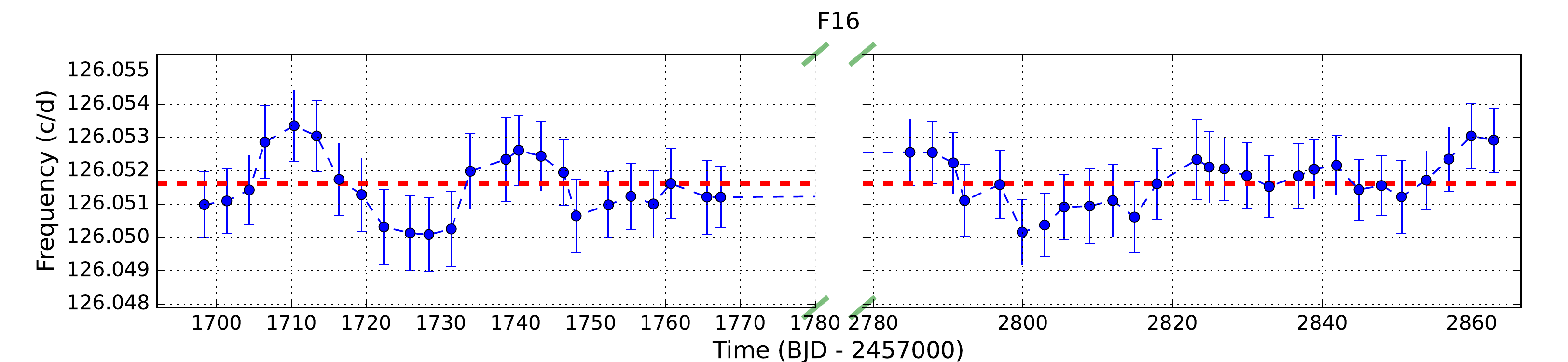}
  \includegraphics[width=0.495\textwidth]{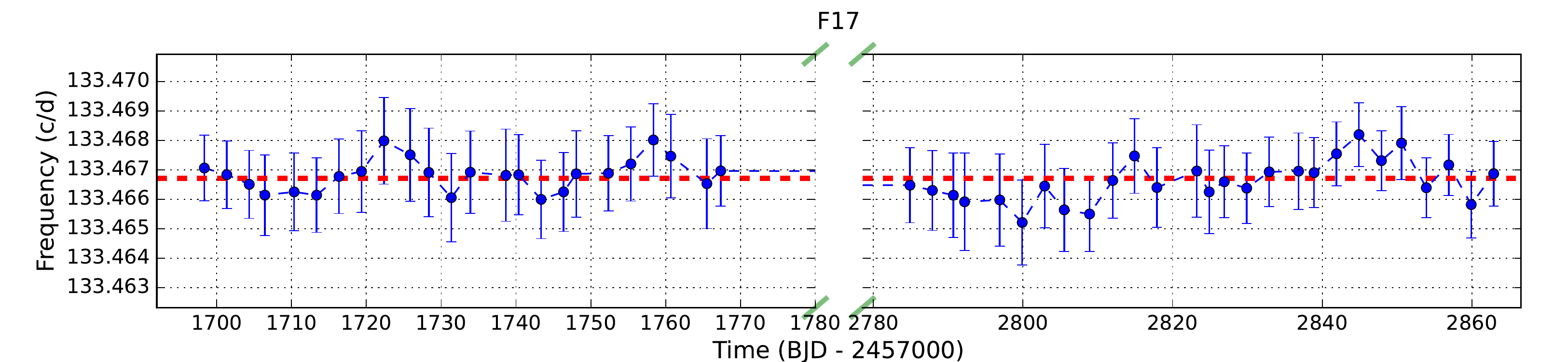}
  \includegraphics[width=0.495\textwidth]{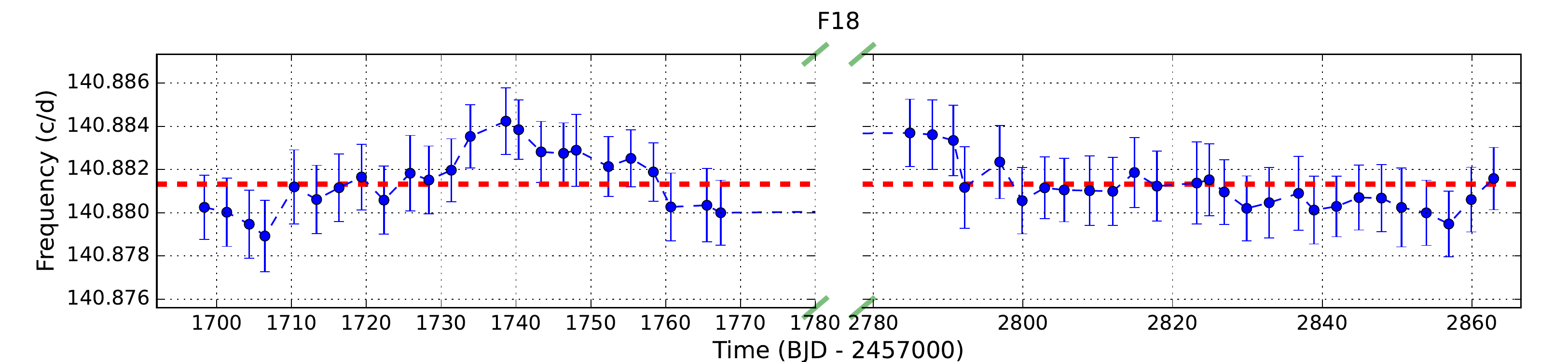}
  \includegraphics[width=0.495\textwidth]{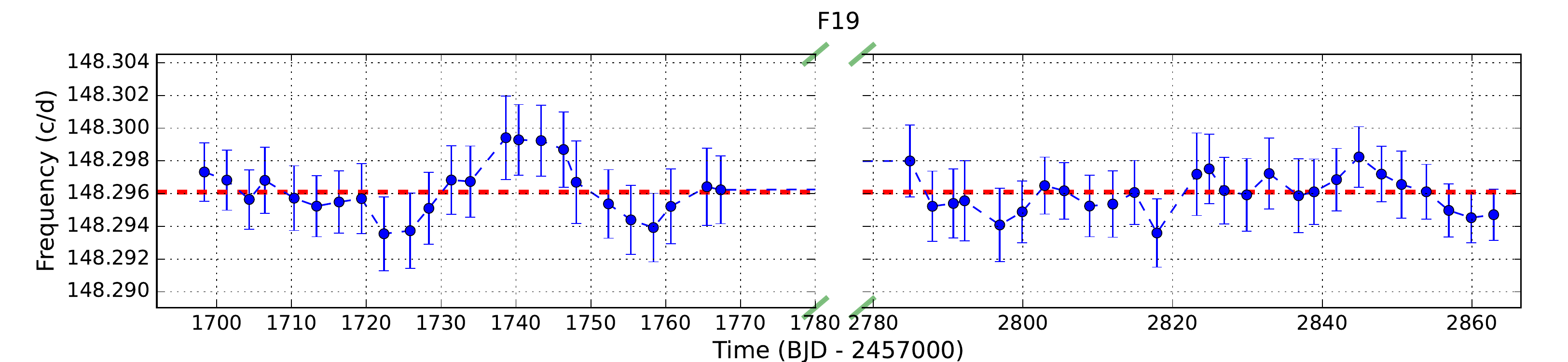}
  \caption{Variations of the frequencies of F0 to F19. The blue dots with errorbars present the frequencies in moving windows, and the red dashed lines present the time-averaged frequencies.}
  \label{fig:var_freq}
\end{figure*}

In order to extract the variation information in the amplitudes and frequencies over time, we used the short-time Fourier transformation to deal with the light curves of DS1 and DS2 separately (see, e.g., \citet{Niu2022} for more details). The prewhitening process in a time window of 30 days was performed when the window was moving from the start to the end time of the data set, with a step of 3 days.
In each step in the prewhitening process, the Lomb-Scargle algorithm \citep{Lomb_Scargle} was used to help find the initial value of frequency with the largest amplitude, and then a nonlinear least-square fitting was performed to get the final values of the frequency, amplitude, and phase \citep{Niu2022}. At last, the fundamental pulsation mode (labeled as F0) and its 19 harmonics (labeled as F1 to F19) were considered in this work due to the signal-to-noise ratio of them in most of the moving time windows. The amplitudes and frequencies of F0 to F19 over time are shown in Figures \ref{fig:var_amp} and \ref{fig:var_freq}.

In this work, the uncertainties of amplitudes ($\sigma_{a}$) are defined as the median value of the amplitudes within a spectral window of 4 $\cd$ equally divided by the frequency peak (similar as that in \citet{Zong2018,Niu2022}), while the uncertainties of the frequencies ($\sigma_{f}$) are estimated following relationship between the amplitude and frequency proposed by \citet{Montgomery1999} and \citet{Aerts2021},
\begin{equation}
  \label{eq:sigma_f}
  \sigma_{f} = \sigma_{a} \cdot \frac{\sqrt{3}}{\pi \cdot  A \cdot T},
\end{equation}
where $A$ is the amplitude, $T$ is the total time baseline employing in the prewhitening process.

\section{Results}

In Figure \ref{fig:var_amp}, it is clear that the amplitudes in DS1 are systematically greater than that in DS2 from F0 to F14, which might be caused by the contamination by neighbouring stars. As the order of the harmonics increases (from F15 to F19), it seems not that obvious and shows more fluctuations. In Figure \ref{fig:var_freq}, it shows that the frequencies are stable within the uncertainties for the fundamental mode and first several harmonics (from F0 to F9), and the situations become complex and more fluctuations emerge as the order of the harmonic increases.

What is more interesting is that the different harmonics show uncorrelated variations both in amplitude and frequency, which is almost impossible to be affected by the contamination by neighbouring stars. For example, the amplitude of F16 increases after 1720 (BJD - 245700), while that of F19 decreases after that time. Meanwhile, the frequency of F16 shows two bumps at about 1710 and 1740 (BJD - 2457000), while that of F19 shows only one bump at about 1740 (BJD - 2457000).

In order to show the uncorrelated variations of the amplitude and frequency of the fundamental pulsation mode and its harmonics, their amplitudes and frequencies are renormalized as $\mathrm{A}i \equiv A_{i}/<A_{i}>$ and $\mathrm{f}i \equiv f_{i}/<f_{i}>$, where $<A_{i}>$ and $<f_{i}>$ denote the time-averaged values of the amplitude $A_{i}$ and frequency $f_{i}$ of F$i$. Meanwhile, the uncertainties of $\mathrm{A}i$ and $\mathrm{f}i$ are obtained by $\sigma_{\mathrm{A}i} \equiv \sigma_{A_{i}}/<A_{i}>$ and $\sigma_{\mathrm{f}i} \equiv  \sigma_{f_{i}}/<f_{i}>$, where $\sigma_{A_{i}}$ and $\sigma_{f_{i}}$ are the uncertainties of the amplitude and frequency of F$i$. The time-averaged normalized amplitude and frequency uncertainties of F0 to F19 ($<\sigma_{\mathrm{A}i}>$ and $<\sigma_{\mathrm{f}i}>$) are listed in Table \ref{tab:norm_uncertanty}.

\begin{table}[!htbp]
\begin{center}
  \caption{Time-averaged uncertainties of A$i$ and f$i$.}
  \label{tab:norm_uncertanty}
 \begin{tabular}{l|cc}
   \hline
   \hline
   {ID}&{$<\sigma_{\mathrm{A}i}>$}&{$<\sigma_{\mathrm{f}i}>$}\\
\hline
   F0 & $1.83 \times 10^{-2}$ &  $4.65 \times 10^{-5}$\\ 
   F1 & $1.81 \times 10^{-2}$ &  $2.31 \times 10^{-5}$\\  
   F2 & $1.82 \times 10^{-2}$ &  $1.54 \times 10^{-5}$\\  
   F3 & $1.83 \times 10^{-2}$ &  $1.17 \times 10^{-5}$\\  
   F4 & $1.86 \times 10^{-2}$ &  $9.44 \times 10^{-6}$\\  
   F5 & $1.88 \times 10^{-2}$ &  $7.96 \times 10^{-6}$\\  
   F6 & $1.90 \times 10^{-2}$ &  $6.89 \times 10^{-6}$\\  
   F7 & $1.98 \times 10^{-2}$ &  $6.29 \times 10^{-6}$\\  
   F8 & $2.12 \times 10^{-2}$ &  $5.99 \times 10^{-6}$\\   
   F9 & $2.23 \times 10^{-2}$ &  $5.67 \times 10^{-6}$\\   
   F10& $2.38 \times 10^{-2}$ &  $5.49 \times 10^{-6}$\\   
   F11& $2.59 \times 10^{-2}$ &  $5.48 \times 10^{-6}$\\   
   F12& $2.99 \times 10^{-2}$ &  $5.83 \times 10^{-6}$\\   
   F13& $3.29 \times 10^{-2}$ &  $5.95 \times 10^{-6}$\\  
   F14& $3.91 \times 10^{-2}$ &  $6.62 \times 10^{-6}$\\  
   F15& $4.54 \times 10^{-2}$ &  $7.20 \times 10^{-6}$\\  
   F16& $5.46 \times 10^{-2}$ &  $8.15 \times 10^{-6}$\\  
   F17& $6.97 \times 10^{-2}$ &  $9.84 \times 10^{-6}$\\  
   F18& $8.30 \times 10^{-2}$ &  $1.11 \times 10^{-5}$\\  
   F19& $1.09 \times 10^{-1}$ &  $1.39 \times 10^{-5}$\\  
   \hline
   \hline
\end{tabular}
\end{center}
\end{table}

\begin{figure*}[!htbp]
  \centering
  \includegraphics[width=0.495\textwidth]{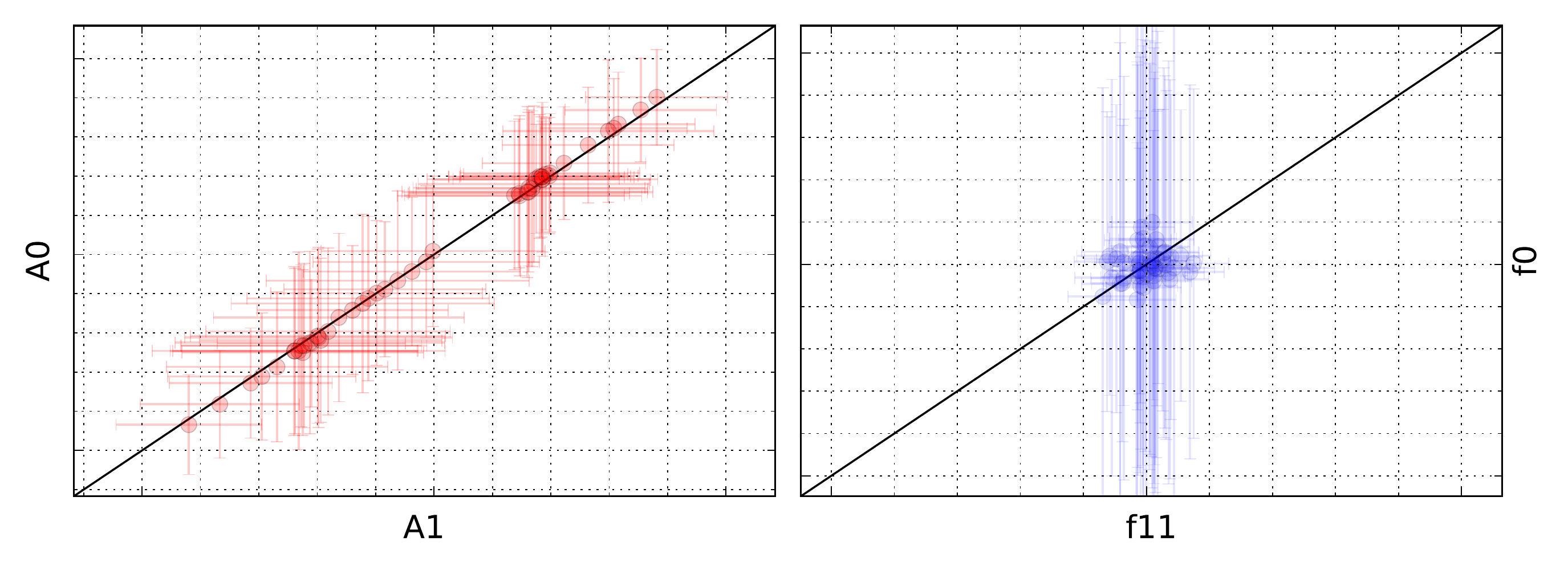}
  \includegraphics[width=0.495\textwidth]{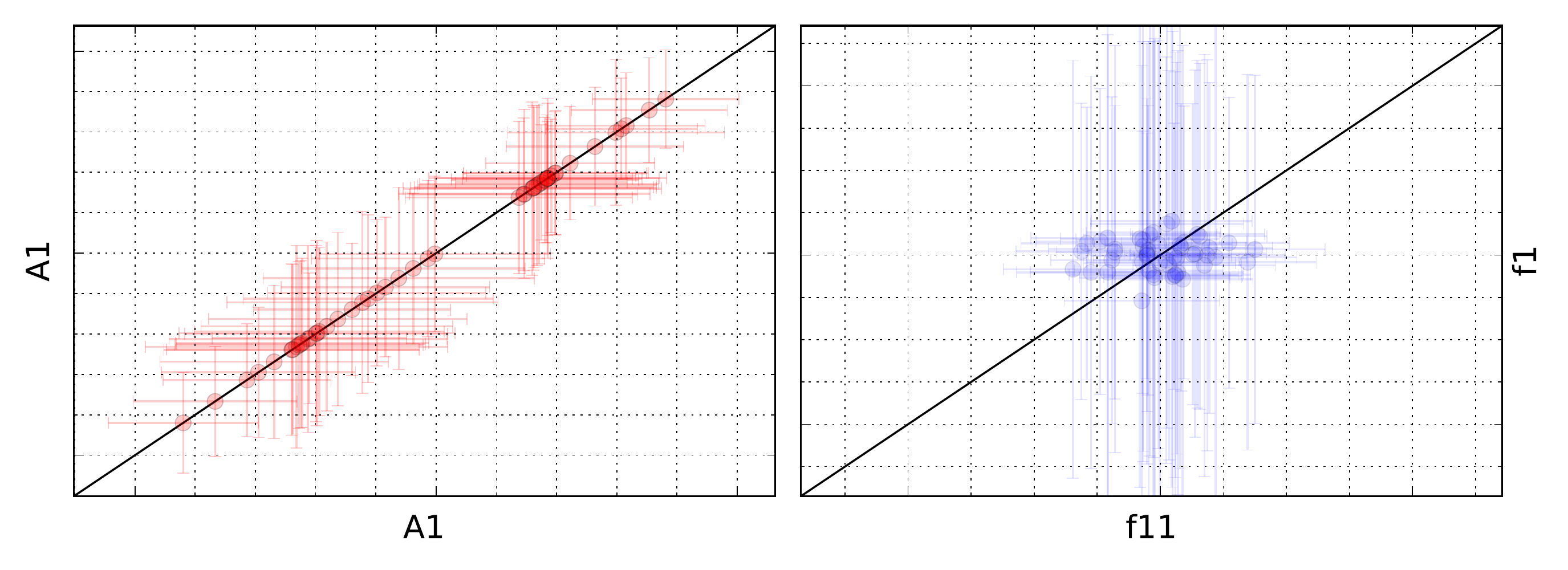}
  \includegraphics[width=0.495\textwidth]{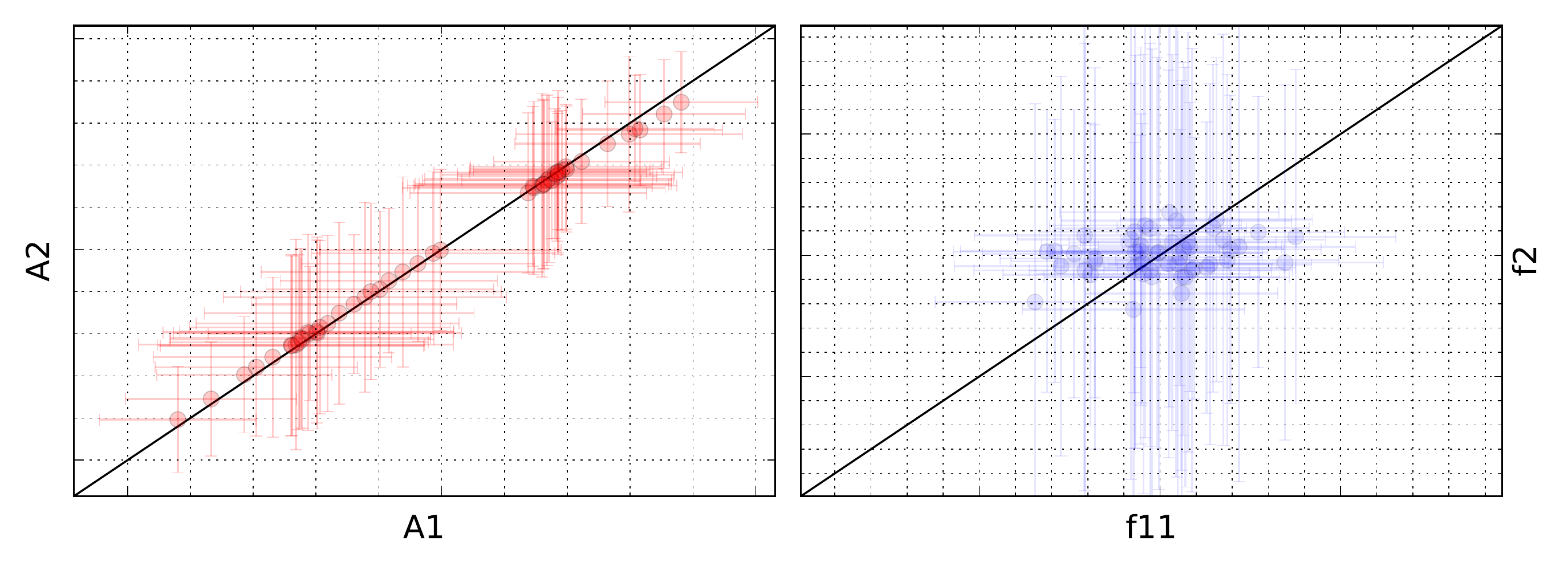}
  \includegraphics[width=0.495\textwidth]{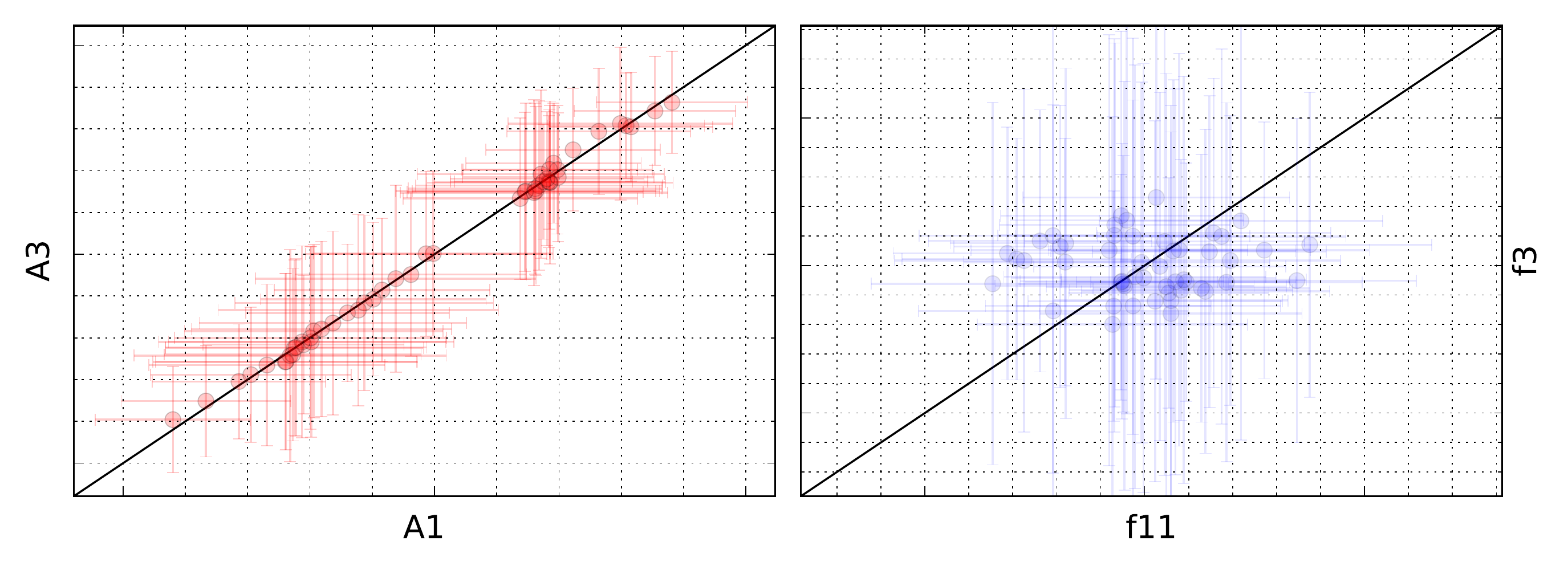}
  \includegraphics[width=0.495\textwidth]{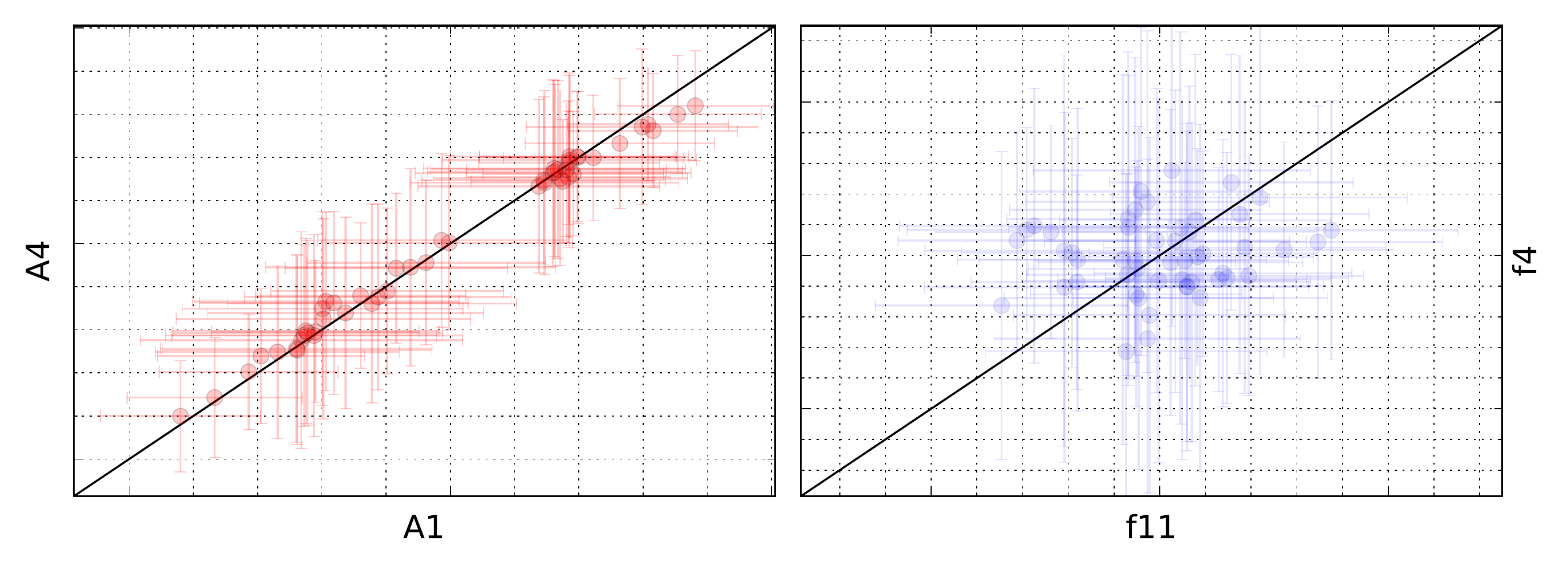}
  \includegraphics[width=0.495\textwidth]{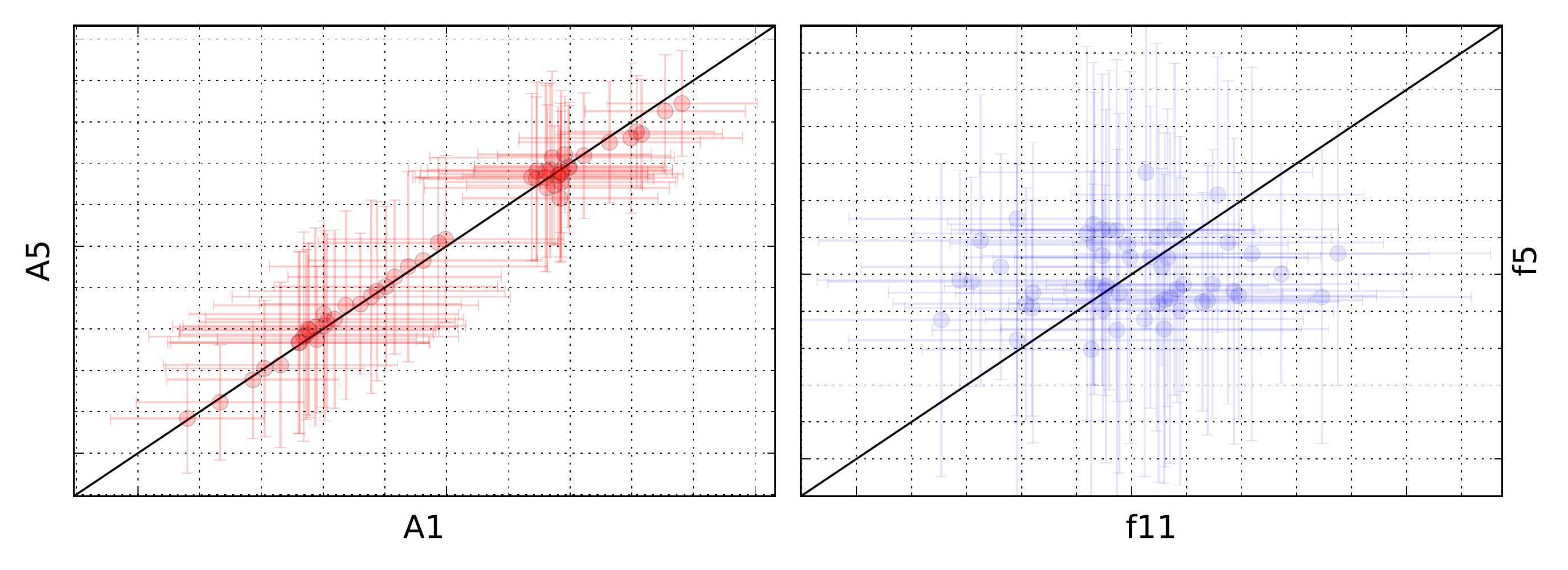}
  \includegraphics[width=0.495\textwidth]{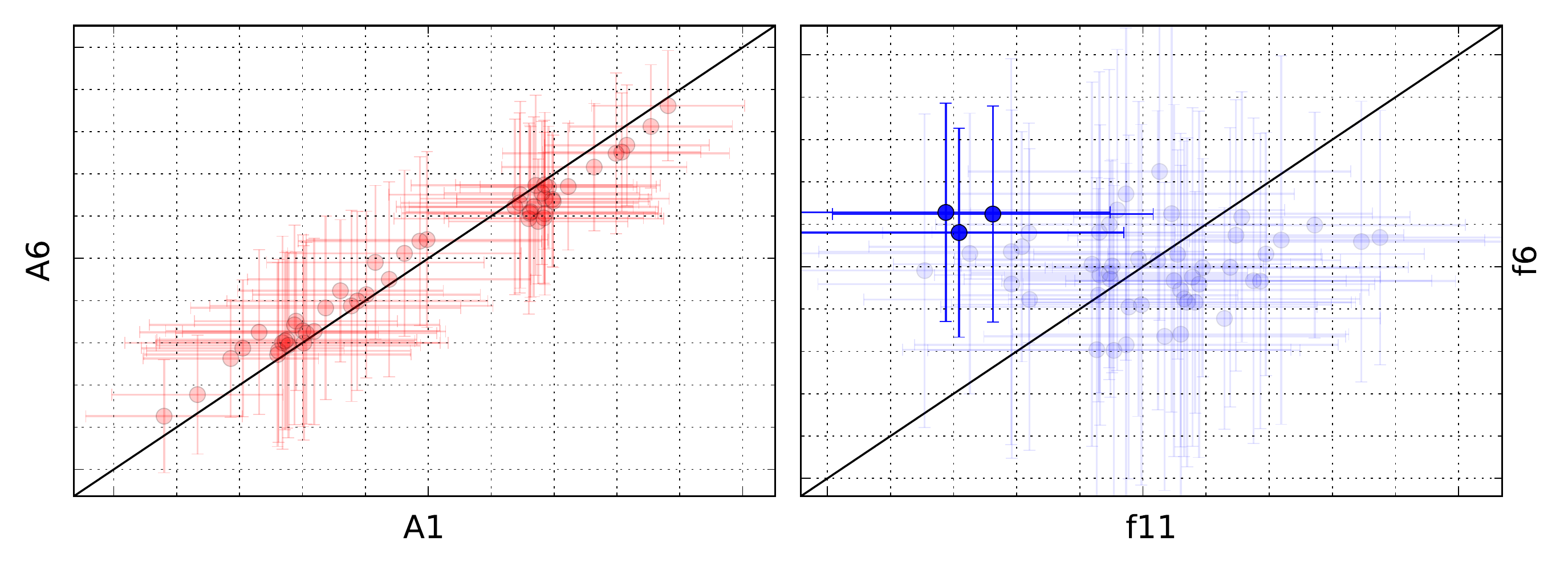}
  \includegraphics[width=0.495\textwidth]{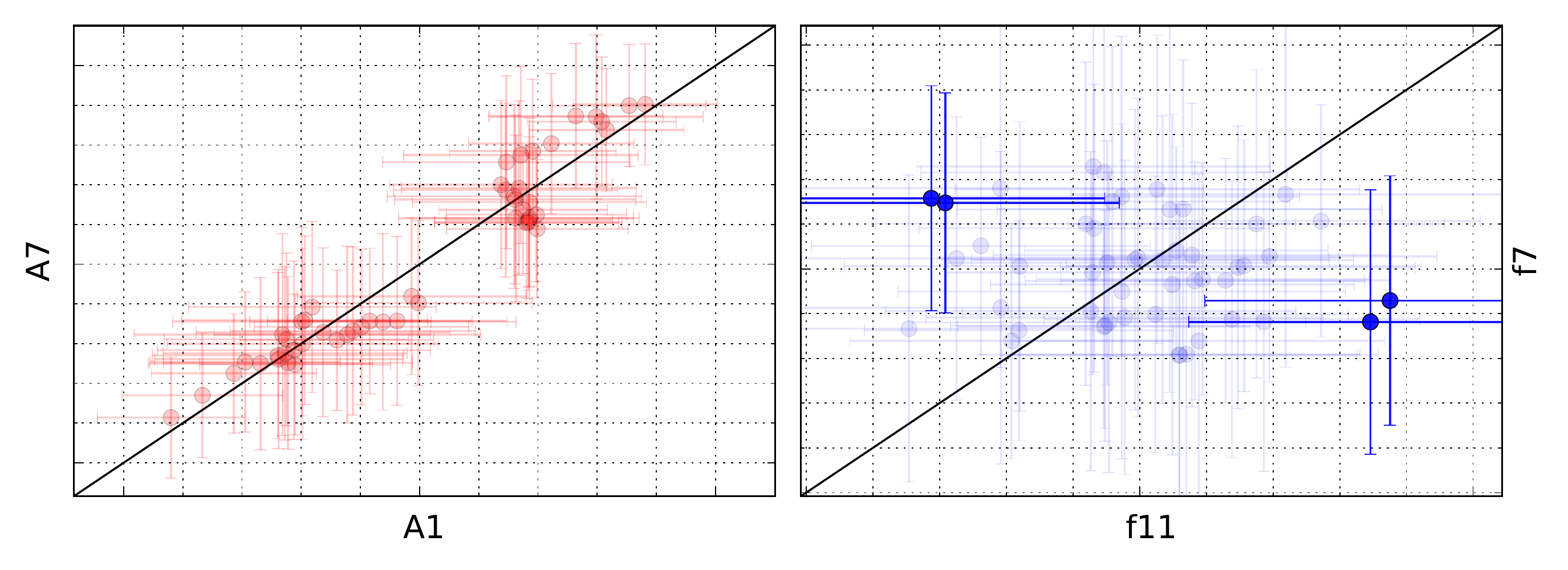}
  \includegraphics[width=0.495\textwidth]{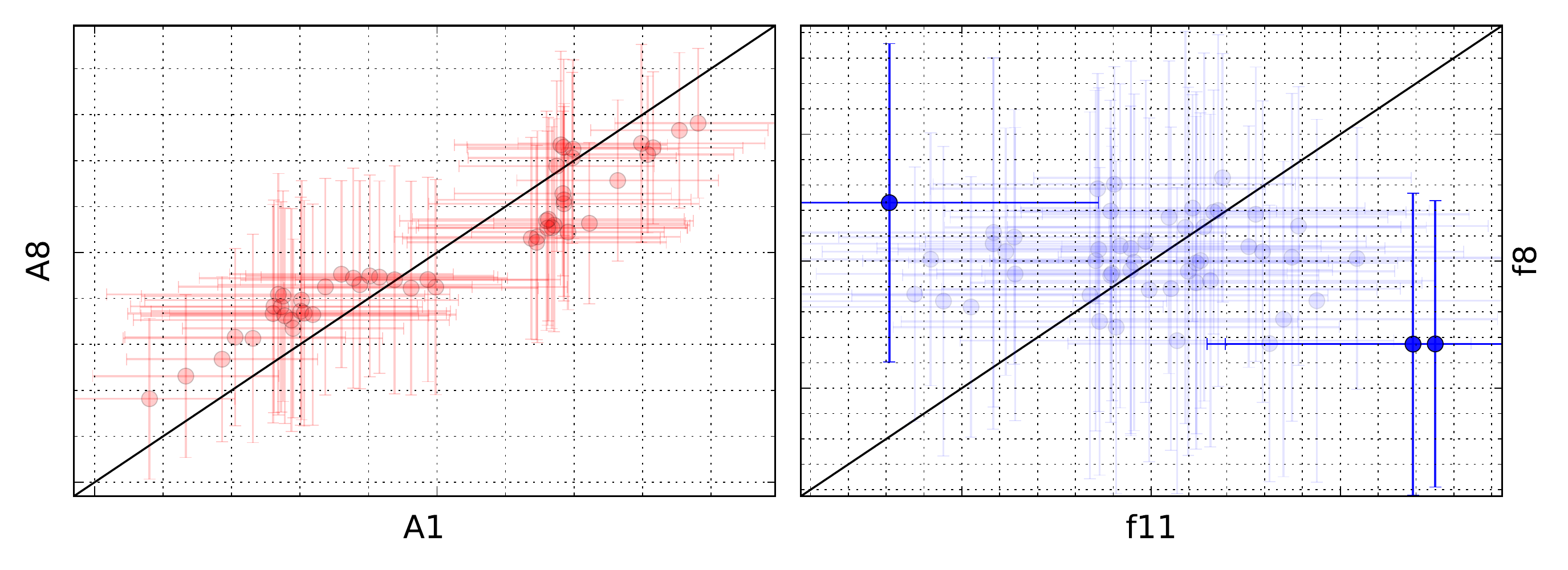}
  \includegraphics[width=0.495\textwidth]{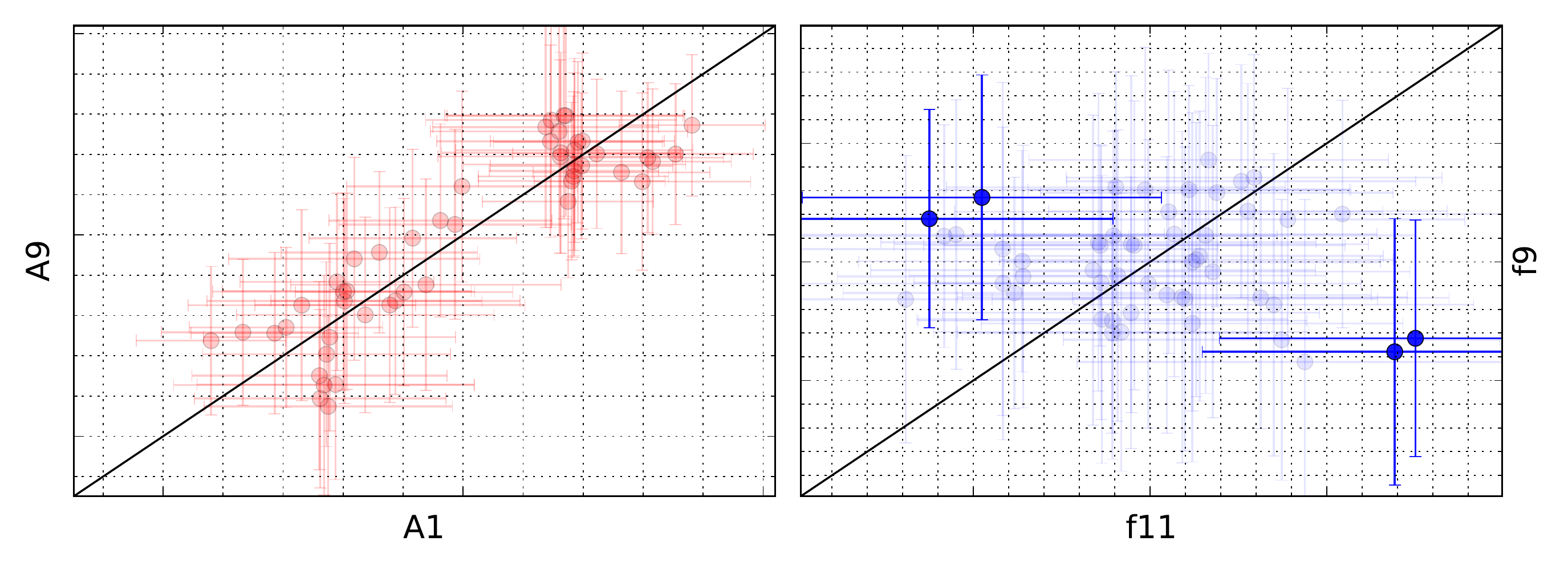}
  \includegraphics[width=0.495\textwidth]{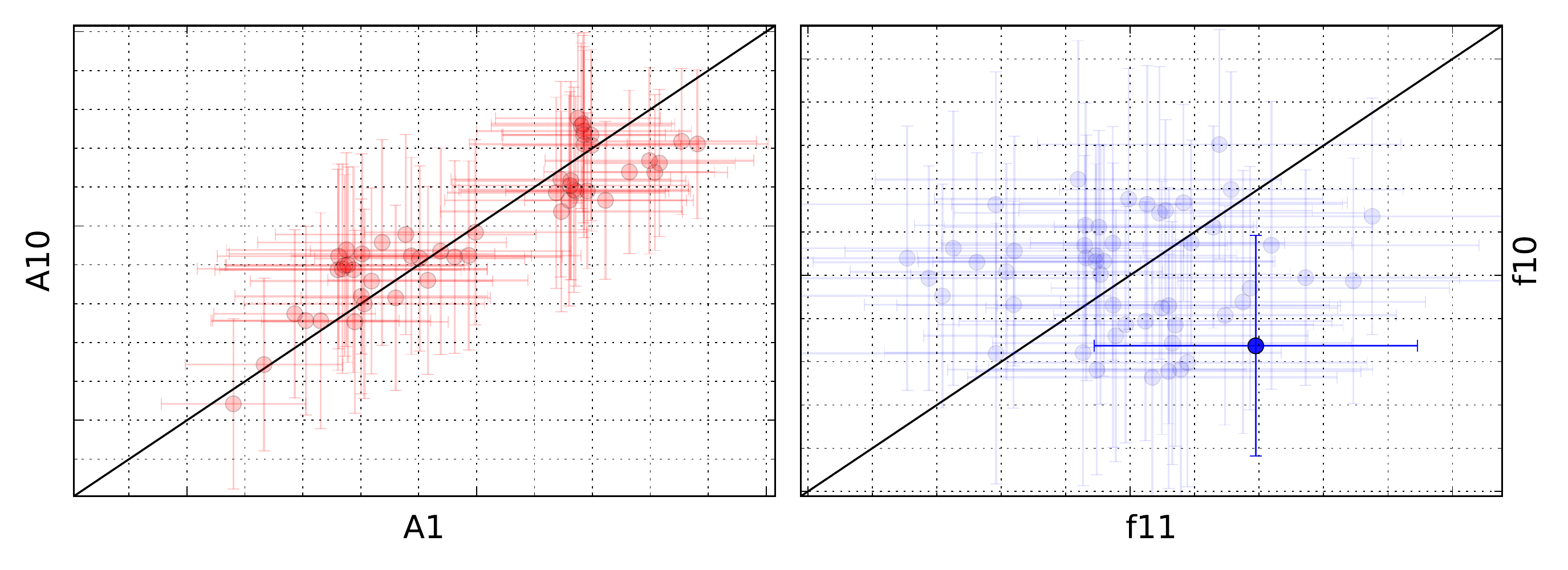}
  \includegraphics[width=0.495\textwidth]{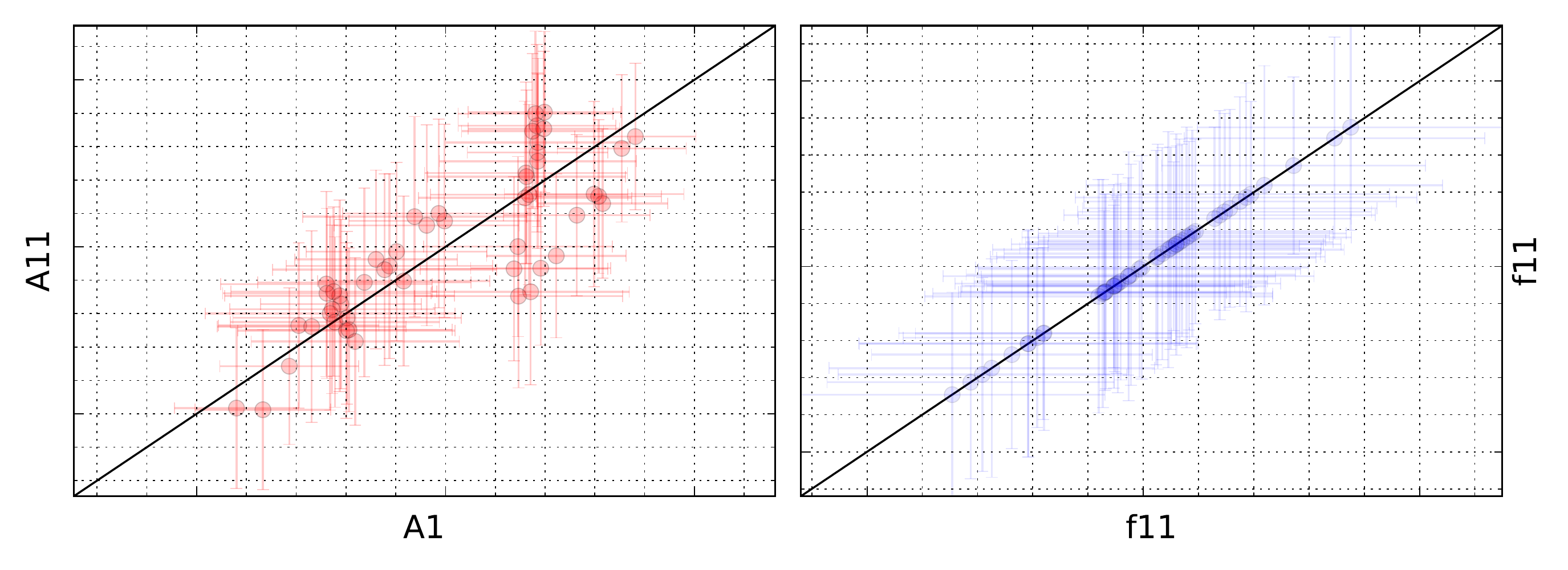}
  \includegraphics[width=0.495\textwidth]{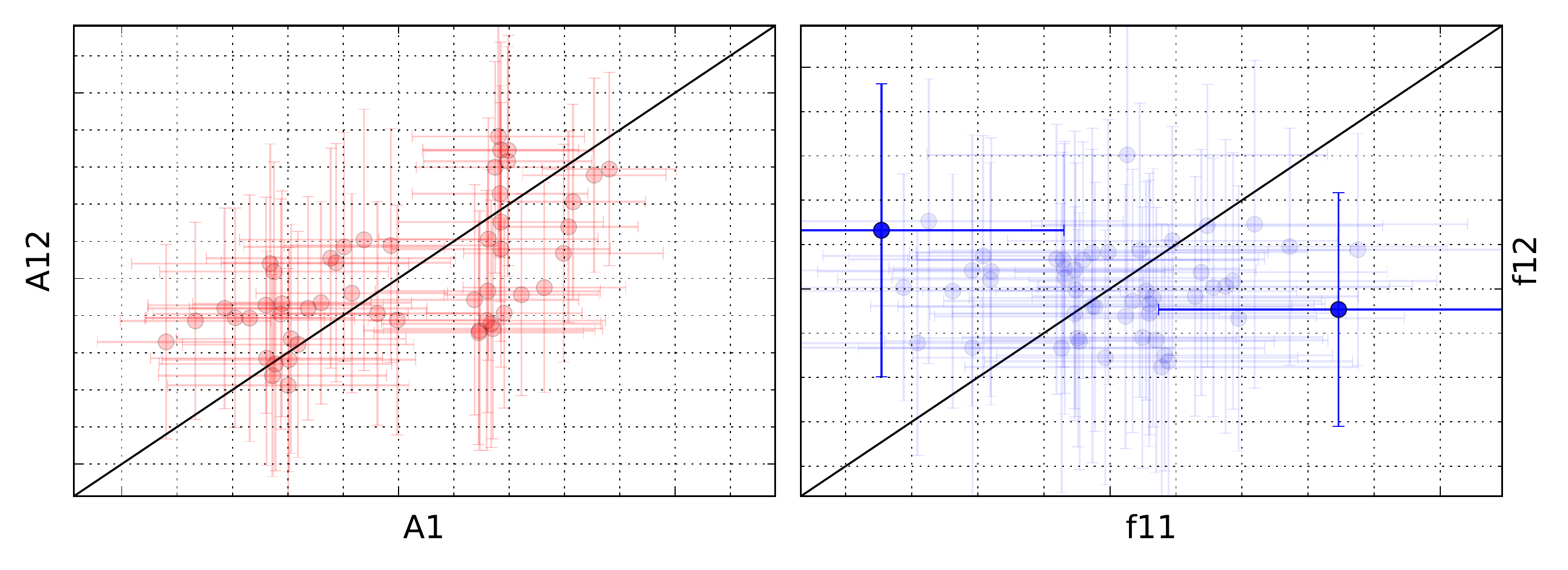}
  \includegraphics[width=0.495\textwidth]{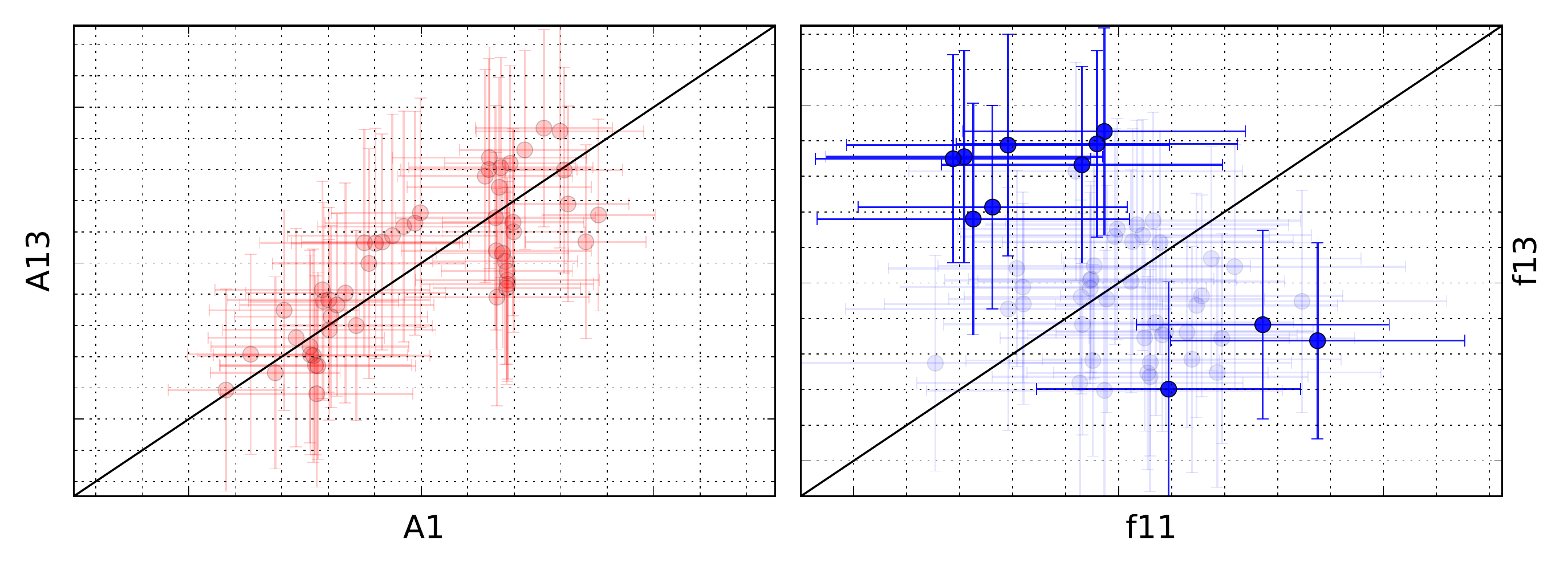}
  \caption{Correlation chart based on F1 (amplitude) and F11 (frequency), Part I.}
  \label{fig:fvsf1111}
\end{figure*}

\begin{figure*}[!htbp]
  \centering
  \includegraphics[width=0.495\textwidth]{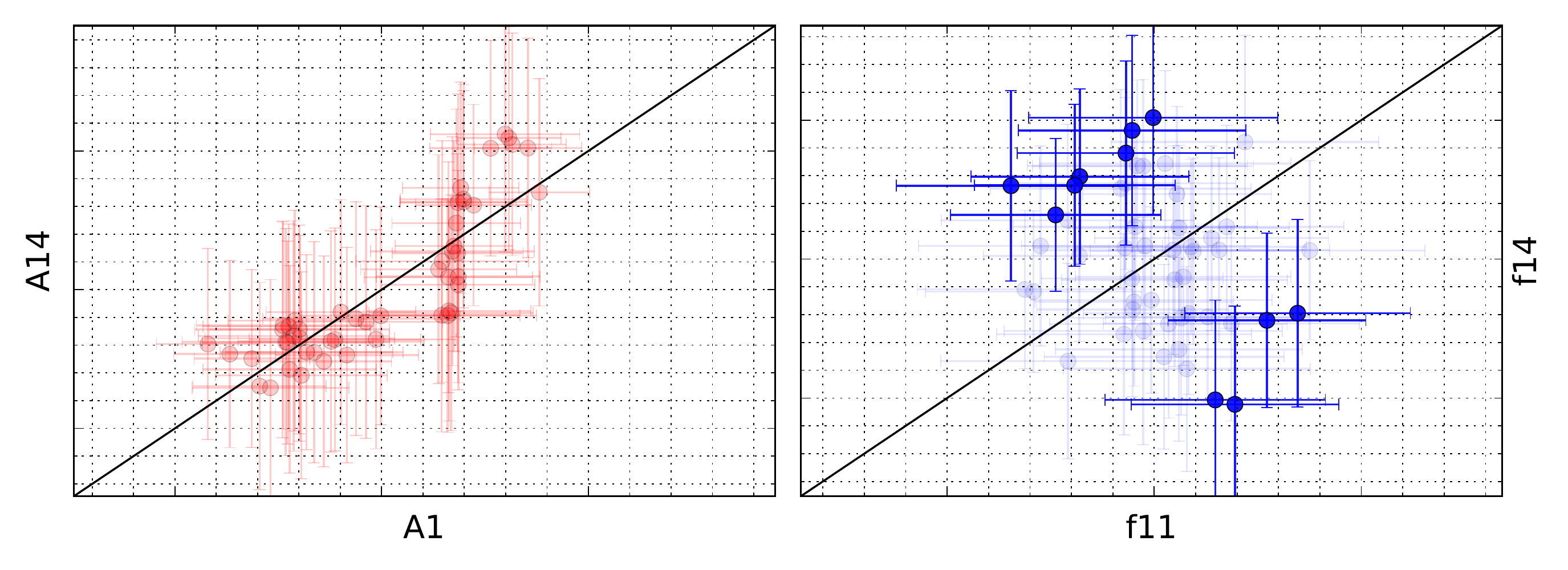}
  \includegraphics[width=0.495\textwidth]{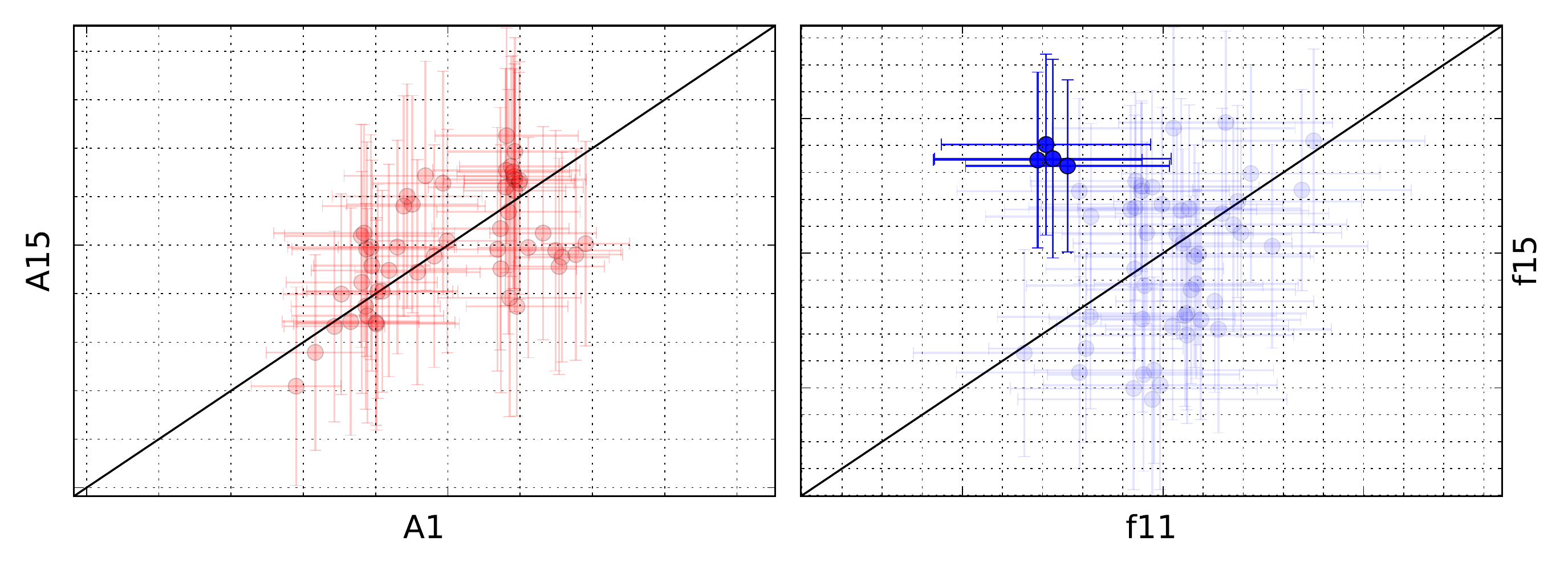}
  \includegraphics[width=0.495\textwidth]{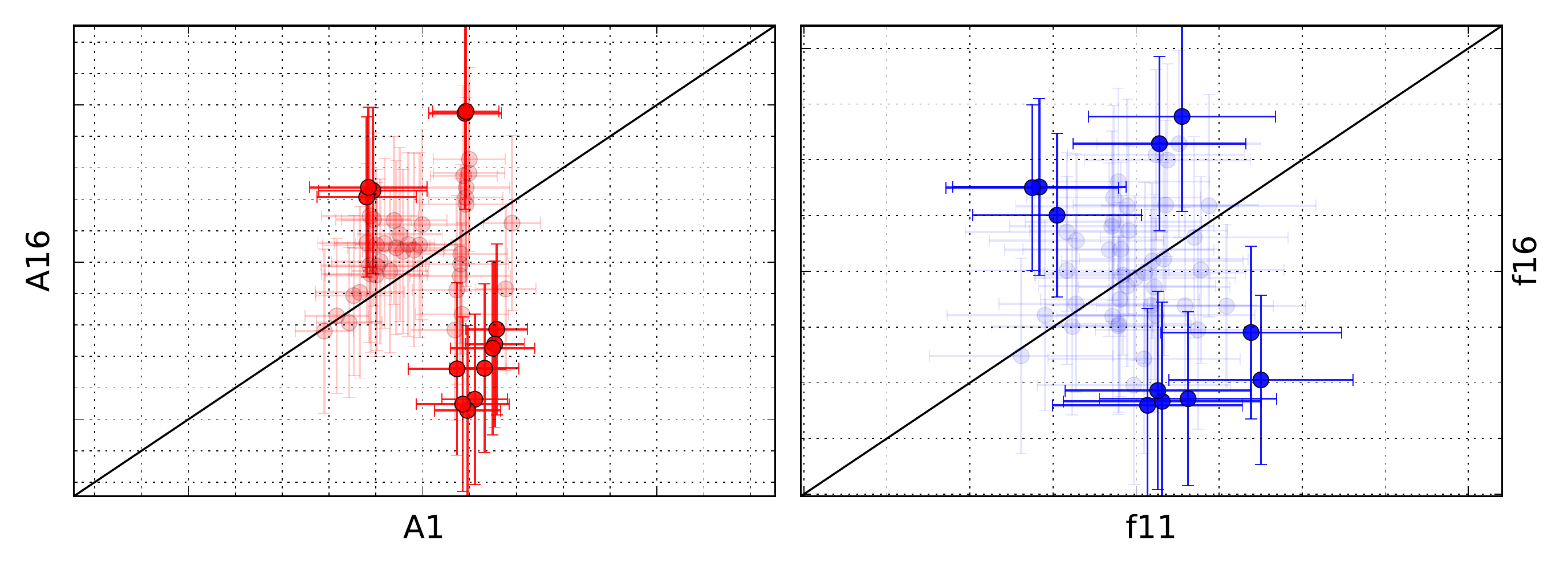}
  \includegraphics[width=0.495\textwidth]{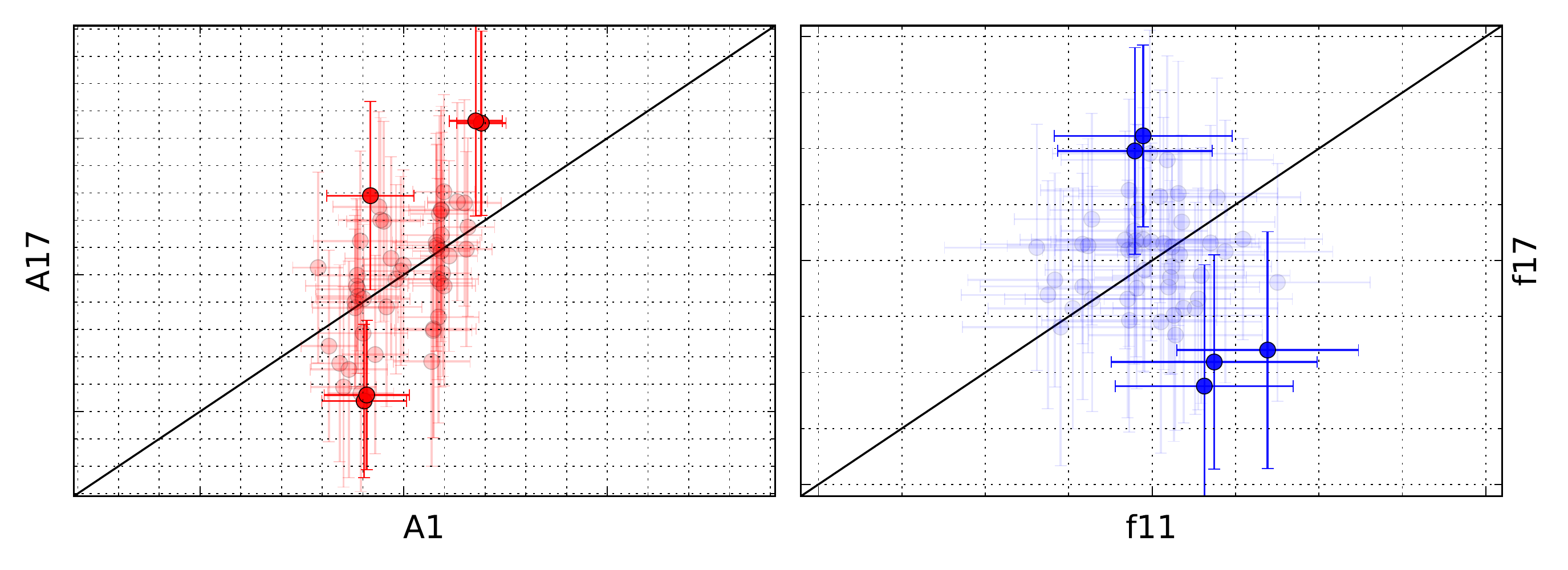}
  \includegraphics[width=0.495\textwidth]{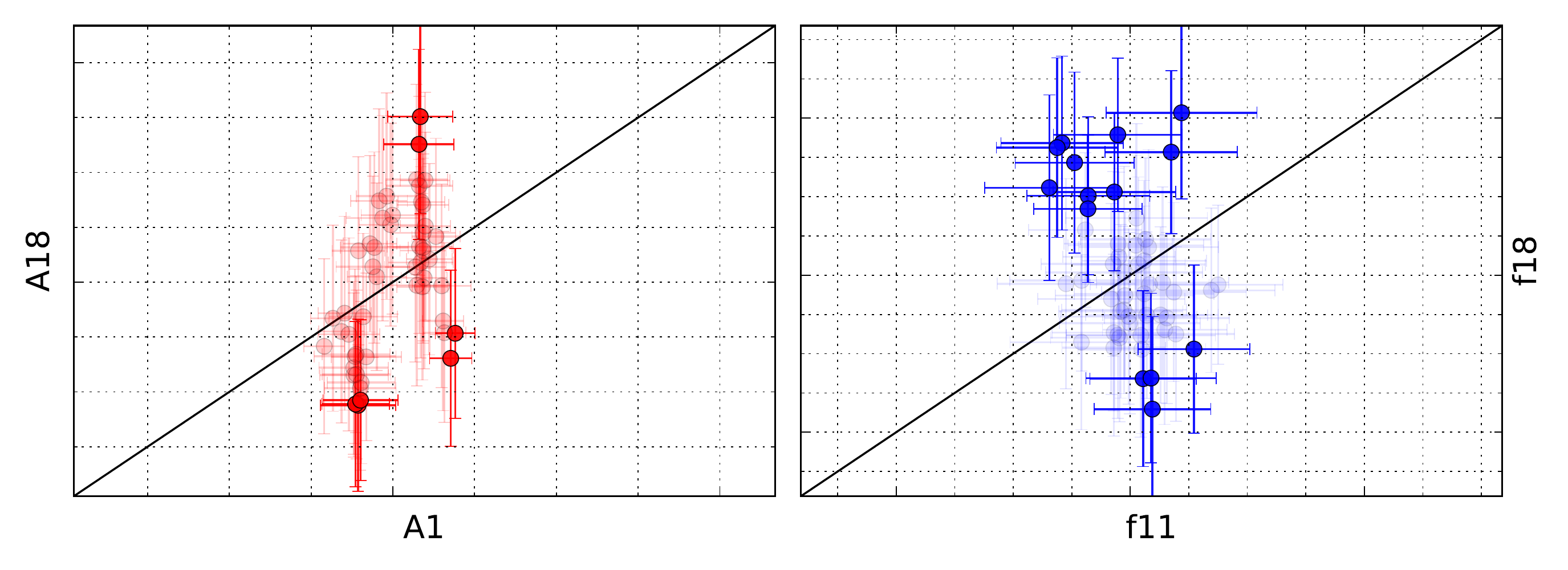}
  \includegraphics[width=0.495\textwidth]{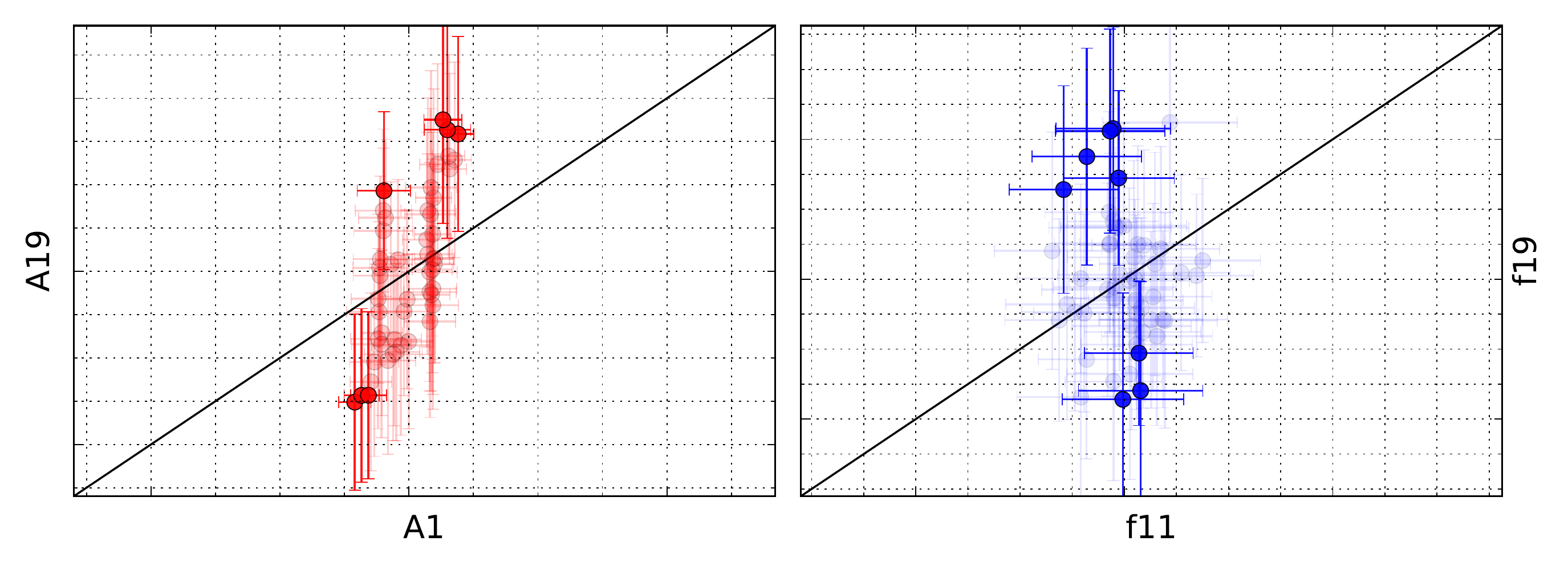}
  \caption{Correlation chart based on F1 (amplitude) and F11 (frequency), Part II.}
  \label{fig:fvsf1112}
\end{figure*}

In order to show the uncorrelated variations of the amplitude and frequency clearly, we choose the amplitudes of F1 (A1) and the frequencies of F11 (f11) as the bases to present the correlation chart (A$i$ vs. A1 and f$i$ vs. f11) in Figures \ref{fig:fvsf1111} and \ref{fig:fvsf1112}, which have the smallest time-averaged uncertainties.

In Figures \ref{fig:fvsf1111} and \ref{fig:fvsf1112}, we treat the points in each of the subfigures obeying the two-dimensional normal distributions $N(\mathrm{A}i,\mathrm{A}1,\sigma_{\mathrm{A}i},\sigma_{\mathrm{A}1},0)$ and $N(\mathrm{f}i,\mathrm{f}11,\sigma_{\mathrm{f}i},\sigma_{\mathrm{f}11},0)$. If the points deviate the diagonals more than $1\sigma$, we plot them with dark red/blue dots (with errorbars) rather than light ones, which helps us to find the significantly deviated points clearly.

For the amplitudes, as the order of the harmonic increases, the tendency to deviate from the diagonals is growing stronger. From A16 to A19, it is clear that some points deviate the diagonal significantly, which indicates that the amplitudes of F16 to F19 show uncorrelated variations comparing with F1.
For the frequency, we cannot find the significantly deviated points from F0 to F5 because of the relatively large uncertainties of $\sigma_{\mathrm{f}i}\ (i=0,1,2,3,4,5)$, although it shows uncorrelated variation tendency with F11. However, from F6 to F19 (except F11 itself), we can confirm that some points deviate the diagonals significantly, which indicates that the frequencies of F6 to F19 show uncorrelated variations comparing with F11.
Consequently, both the amplitudes and frequencies of the harmonics show uncorrelated variations, which is a challenge to our common perception that the harmonics should mimic the behaviors of their parent pulsation mode. There must be some unrevealed mechanisms hidden inside.

\section{Discussions and Conclusions}

In history, the period/frequency variation of the dominating pulsation mode of some kinds of pulsating stars can be obtained via the accumulation of the TML lasts for several decades by $O-C$ method (such as Cepheid \citep{Csornyei2022,YacobA2022}, RR Lyrae stars \citep{Li2018}, HADS \citep{Xue2022}, and SX Phe stars \citep{Xue2020}). An interesting but often overlooked detail is that it always exists dispersions in the $O-C$ diagrams, which are always one or more orders of magnitude larger than the TML uncertainties \citep{Csornyei2022,YacobA2022,Li2018,Xue2020,Xue2022}. This phenomenon not only happens for multi-mode pulsating stars, but also for the single-mode ones. In the former case, we can ascribe it to the perturbations from the non-dominating pulsation modes to the dominating one. However, in the latter case, the $O-C$ dispersions (about $0.001-0.005\ \mathrm{d}$ for XX Cyg, see in Figure 6 of \citet{Yang2012}) should not exceed the level of TML uncertainties (about $0.0001\ \mathrm{d}$ for XX Cyg, see in Table 3 of \citet{Yang2012}) too much, unless the apparently stable pulsation mode is actually not stable.

As that has been revealed in this work, the $O-C$ dispersions in these single-mode pulsating stars can also be represented by the uncorrelated amplitude and frequency variations of the harmonics. 
As a result, the uncorrelated amplitude and frequency variation should be a common phenomenon in pulsating stars which show harmonics in their periodograms, not only for the single-mode ones, but also for the multi-modes ones.

What are the origins of these uncorrelated variations?
It could come from the fundamental pulsation mode or from the factors outside it.
If we assume it comes from the fundamental mode, the physical origin might be the highly nonlinear interactions between the high-order harmonics in the outer layers (e.g., the dynamical interaction between a multi-shock structure and an outflowing wind in the coronal structure \citep{Chadid2014}).
If we assume it comes from the factors beyond the fundamental mode, it could be caused by the perturbations from the hidden high-order p modes in the outer layers.
Of course, these uncorrelated variations could be the results of a combination of these factors.

Another aspect of interest is that what is the time scale of the uncorrelated variation? A relatively definite time scale would tell us which physical process it closely related to, and a stochastic time scale could be caused by some stochastic mechanisms and related to the stochastic excited modes.
This issue should be studied in depth in future based on the accumulation of the high-precision time-series photometric data, which could shed light on some hidden corners of asteroseismology.

\section*{Acknowledgements}
%\begin{acknowledgements}
J.S.N. acknowledges support from the National Natural Science Foundation of China (NSFC) (No. 12005124 and No. 12147215). H.F.X. acknowledges support from the Scientific and Technological Innovation Programs of Higher Education Institutions in Shanxi (STIP) (No. 2020L0528) and the Applied Basic Research Programs of Natural Science Foundation of Shanxi Province (No. 202103021223320).
%\end{acknowledgements}
All the {\it TESS} data used in this paper can be found in MAST: \dataset[10.17909/t9-nmc8-f686]{http://dx.doi.org/10.17909/t9-nmc8-f686}.
\software{Astropy \citep{astropy}}

%%\clearpage

%%\bibliography{ref}

\label{lastpage}

\end{CJK*}
\end{document}